\documentclass[%
 aip,
 amsmath,amssymb,
 reprint,%
]{revtex4-1}

\usepackage{amsmath, amssymb, bm}
\usepackage{graphicx, subcaption}
\usepackage{tabularx}
\usepackage{algorithm}
\usepackage{algpseudocode}

\usepackage{dcolumn}

\usepackage[utf8]{inputenc}
\usepackage[T1]{fontenc}
\usepackage{mathptmx}
\usepackage{etoolbox}

\makeatletter
\def\@email#1#2{%
 \endgroup
 \patchcmd{\titleblock@produce}
  {\frontmatter@RRAPformat}
  {\frontmatter@RRAPformat{\produce@RRAP{*#1\href{mailto:#2}{#2}}}\frontmatter@RRAPformat}
  {}{}
}%
\makeatother
\begin{document}

\preprint{AIP/123-QED}

\title[Hybrid deep learning-based phase diversity method for wavefront reconstruction]{Hybrid deep learning-based phase diversity method for wavefront reconstruction}
\author{Y. Rodimkov}
\email{rodimkov@itmm.unn.ru}
\affiliation{Lobachevskii State University of Nizhni Novgorod, pr. Gagarina 23, Nizhni Novgorod, 603022 Russia}
\author{A. Kotov}
\affiliation{Institute of Applied Physics, Russian Academy of Sciences, Nizhny Novgorod 603951, Russia}
\author{K. Burdonov}
\affiliation{Institute of Applied Physics, Russian Academy of Sciences, Nizhny Novgorod 603951, Russia}
\author{S. Perevalov}
\affiliation{Institute of Applied Physics, Russian Academy of Sciences, Nizhny Novgorod 603951, Russia}
\author{V. Volokitin}
\affiliation{Lobachevskii State University of Nizhni Novgorod, pr. Gagarina 23, Nizhni Novgorod, 603022 Russia}
\author{I. Meyerov}
\affiliation{Lobachevskii State University of Nizhni Novgorod, pr. Gagarina 23, Nizhni Novgorod, 603022 Russia}

\author{A. Soloviev}
\affiliation{Institute of Applied Physics, Russian Academy of Sciences, Nizhny Novgorod 603951, Russia}

\date{\today}

\begin{abstract}

The efficiency of high-power laser systems is limited by wavefront distortions in the beam, particularly non-common path aberrations, which reduce the peak intensity at the focal plane. Compensating for these aberrations requires the calibration of the adaptive optics system. Conventional calibration methods rely on a time-consuming iterative optimization that is highly sensitive to initial conditions. While deep learning-based models offer high speed, they often demonstrate insufficient accuracy. In this work, we present a hybrid wavefront reconstruction method that combines a convolutional neural network to generate an initial estimate of the wavefront distortions, with the L-BFGS (Limited-memory Broyden–Fletcher–Goldfarb–Shanno) algorithm for its subsequent refinement. In numerical simulations, the method achieved an efficiency of $\sim 0.99$ in 80\% of the cases for a root-mean-square (RMS) of wavefront distortions ranging from 0 to $1.3\lambda$. In a physical experiment, for initial wavefront distortions with RMS values from 0.15 to $0.6\lambda$, the method achieved an efficiency of $\sim 0.75$. As a result, focusing with a Strehl ratio of $0.96 \pm 0.02$ was attained within 2 to 4 iterations of the algorithm, confirming the applicability of the method for the fast and accurate calibration of adaptive optics systems under real experimental conditions.

\end{abstract}

\maketitle

\section{Introduction}

The primary factors limiting the performance of optical systems are static and dynamic aberrations, which degrade the image quality and reduce the peak intensity at the focal plane. The application of adaptive optics systems (AOS) to compensate for these distortions and achieve diffraction-limited performance is critical for a wide range of applications, including astronomical telescopes \cite{dean2006phase,fienup1993hubble}, microscopy \cite{jin2020wavefront,booth2014adaptive}, ophthalmology \cite{sulai2014non}, optical communication systems \cite{bennet2018free, tyson1996adaptive, li2017bp}, and laser systems \cite{soloviev2024research, yoon2019achieving, pirozhkov2017approaching}. 

\begin{figure}
    \centering
    \includegraphics[width=0.5\textwidth]{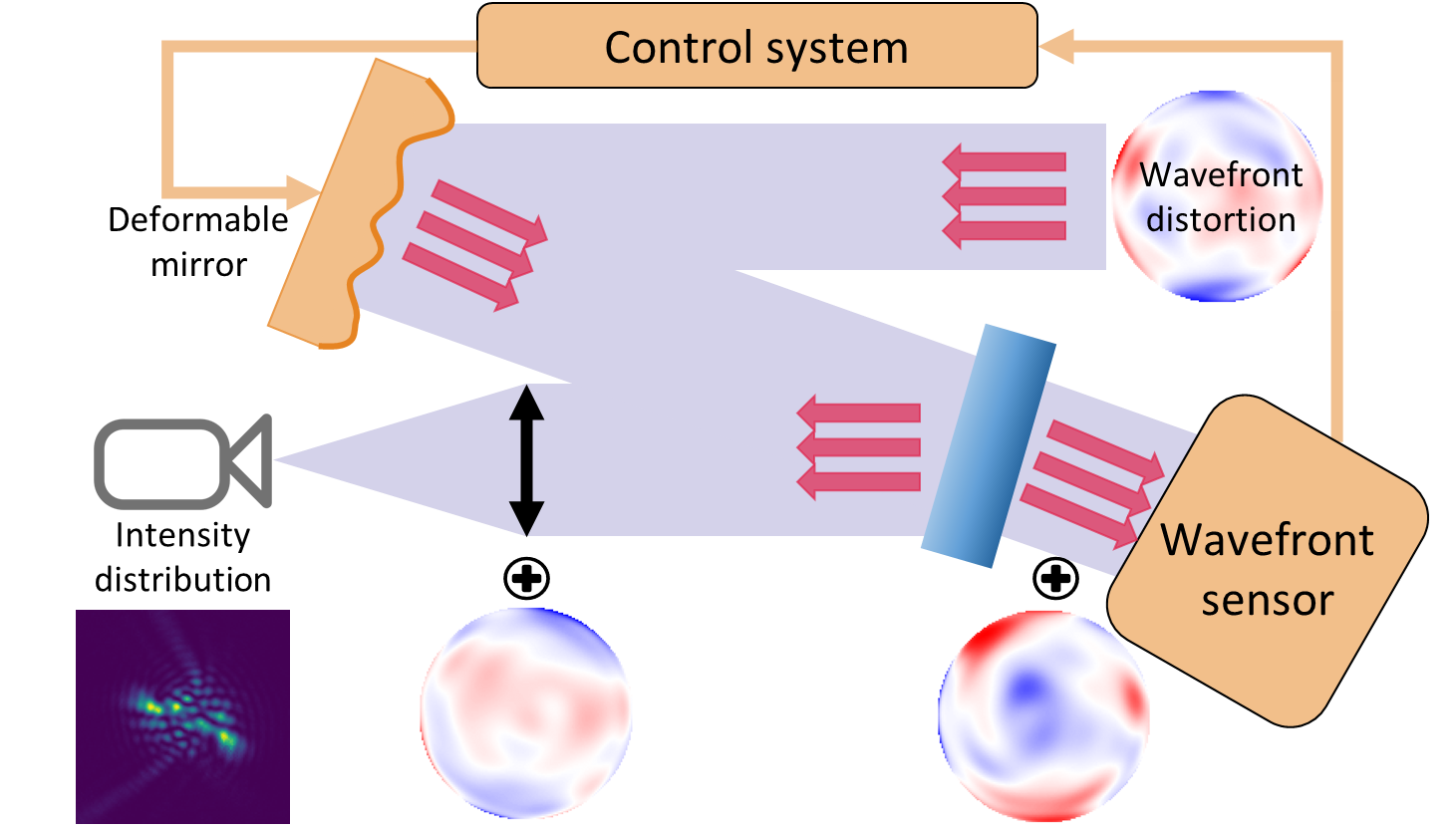}
    \caption{\label{fig:scheme} Schematic of the physical experiment and the operation of the adaptive optics system.}
\end{figure}

A typical AOS configuration includes a wavefront sensor (WFS), a phase-correcting element, such as a deformable mirror (DM), and a control system. In high-power laser facilities, wavefront correction is typically implemented using an AOS, the schematic diagram of which is shown in Fig. \ref{fig:scheme}. The system operates as follows. The laser beam is reflected from the deformable mirror and directed to a partially transparent mirror with a reflectivity of approximately 99\%. The main part of the beam reflected from the partially transparent mirror enters the focusing channel and is focused at the main focal point of the system. The part of the beam that passes through the partially transparent mirror enters the diagnostic channel, where it is used to measure the wavefront. However, this approach suffers from a significant drawback due to non-common path aberrations (NCPA) \cite{sauvage2007calibration}, which are non-identical wavefront distortions acquired by the beam in the diagnostic and focusing channels. Under these conditions, even if a perfectly flat wavefront is achieved at the WFS plane, the resulting intensity distribution at the focal plane will not reach the diffraction limit. 

The presence of NCPA necessitates an AOS calibration procedure to determine the reference wavefront for the WFS that corresponds to optimal focusing at the target. Essentially, this reference wavefront is the conjugate of the non-common path wavefront distortions. In practice, any changes in the focusing or diagnostic channels require recalibration, so it is usually performed before each experimental session, as well as during experiments.

Existing methods for determining the reference wavefront can be divided into three main categories, each with inherent limitations. (i) Phase retrieval methods, based on the intensity distributions in the near and far fields \cite{gerchberg1972practical,gonsalves1982phase}, rely on the iterative propagation of the optical field between the near- and far-field planes via the Fourier transform, replacing the computed distribution of the electric field amplitude with the one derived from the measured distribution at each step. Their drawbacks include high computational complexity, sensitivity to noise, instability under complex wavefront distortions, and susceptibility to artifacts. (ii) Iterative methods \cite{piatrou2007beaconless, yang2007intracavity} for optimizing the intensity distribution at the focal plane within the parameter space of the wavefront distortions require a large number of iterations, often converge to local extrema, and yield results that are highly dependent on initial conditions and optimization parameters. (iii) Recently, neural network-based models \cite{paine2018machine, orban2021focal, kotov2025retrieval} have gained widespread adoption due to their rapid inference speeds. However, their application is limited by insufficient accuracy in reconstructing higher-order wavefront distortions and problems with the generalization ability of models.

To overcome these limitations, we developed hybrid method for reconstruction of wavefront distortions that combines the advantages of iterative optimization and deep learning. The method employs a phase diversity technique, wherein a convolutional neural network generates an initial estimate of the wavefront distortions based on the intensity distributions at the focus and defocused planes \cite{kotov2025retrieval}. This initial prediction is subsequently refined using the L-BFGS (Limited-memory Broyden–Fletcher–Goldfarb–Shanno) \cite{liu1989limited} optimization algorithm. This paper presents both numerical and experimental evaluations of the performance of the method. In numerical simulations, a correction efficiency approaching unity is demonstrated using up to 100 Zernike modes \cite{wyant1992basic} for an RMS of wavefront distortions ranging from 0 to $1.3\lambda$. Experimentally, we demonstrate the ability to achieve a Strehl ratio of approximately 0.96 within 2 to 4 iterations of the algorithm for initial wavefront distortions with RMS values up to $0.6\lambda$, which corresponds to a Strehl ratio below 0.1. 

The paper is organized as follows. Section \ref{sec:rel_work} presents an overview of existing methods for the reconstruction of wavefront distortions and the calibration of adaptive systems. Section \ref{sec:method} describes the training data generation procedure and the hybrid method. Section \ref{sec:result} discusses the performance of the method on numerical and experimental data. Section \ref{sec:conclusion} summarizes the main findings of the study.

\section{\label{sec:rel_work}Related work} 

\subsection{Mirror control}

Traditional AOS calibration methods belong to the class of sensorless techniques, which operate without a WFS and are based on iterative algorithms for optimizing the intensity distribution at the focal plane. For the direct control of the DM surface, various optimization algorithms are applied, including hill climbing \cite{sheldakova2004genetic, yang200719}, genetic algorithms \cite{sheldakova2004genetic, yang200719, poland2008evaluation}, stochastic gradient descent \cite{piatrou2007beaconless, cao2014stochastic}, and simulated annealing \cite{el2005adaptive, zommer2006simulated}. In recent years, reinforcement learning methods \cite{ke2019self, landman2021self} have gained widespread adoption, enabling a reduction in the number of iterations required and accelerating the overall calibration process. 

Rather than directly controlling the mirror surface, some approaches employ a mapping between the mirror actuators and a set of basis functions describing the wavefront distortions. Within this formulation, both classical optimization algorithms \cite{kotov2021adaptive, yang2007intracavity, ma2011full, liu2013hill} and modern reinforcement learning techniques \cite{durech2021wavefront, gutierrez2024image} have been explored. A key advantage of this approach is a significant reduction in the dimensionality of the search space compared to a direct actuator-level optimization.

\subsection{Wavefront distortion reconstruction}

An alternative approach to AOS calibration relies on wavefront reconstruction using the phase diversity method proposed by Gonsalves \cite{gonsalves1982phase}. This method enables the unambiguous determination of wavefront distortions from the near-field intensity distribution combined with two intensity measurements taken at the focal and defocused planes. Subsequent studies have compared various optimization algorithms for reconstructing both the Zernike coefficients and the pixel-wise wavefront distortions \cite{fienup1993phase}, adapting these algorithms to account for Gaussian and Poisson noise \cite{paxman1992joint}. Additionally, approaches combining global and local optimization algorithms have been proposed to improve the reconstruction accuracy \cite{zhang2016hybrid, zhou2021robust}. 

However, classical phase retrieval and sensorless mirror control methods exhibit several limitations. Specifically, they are computationally expensive, are prone to stagnation in local extrema, and yield results that are highly sensitive to hyperparameter tuning. Moreover, in the presence of dynamic aberrations, these  distortions can severely interfere with the iterative wavefront correction process \cite{kotov2021adaptive}.

Various convolutional neural network architectures have been investigated and adapted for estimating the Zernike mode coefficients, including AlexNet \cite{jin2018machine}, ResNet \cite{li2022prediction, andersen2020image, orban2021focal, vanberg2019machine}, EfficientNet \cite{li2022prediction}, and Inception \cite{paine2018machine, andersen2020image, vanberg2019machine}. For direct phase retrieval tasks, image-to-image architectures such as De-VGG \cite{guo2019improved}, U-Net \cite{orban2021focal}, and UNet++ \cite{vanberg2019machine} have been successfully applied. These methods demonstrate comparable accuracy \cite{orban2021focal}, though in some cases, architectures like U-Net and UNet++ outperform networks designed solely for Zernike coefficient regression \cite{vanberg2019machine}.

Hybrid methods \cite{paine2018smart, gu2020algorithm, zhou2023generalization, wang2025improved} aim to surpass both the convergence speed of classical local optimization algorithms and the accuracy of neural networks. In these approaches, a deep learning model provides a rapid initial estimate of the wavefront distortions, which is subsequently refined using local optimization algorithms. Nevertheless, existing studies typically focus on a limited number of Zernike modes or lack comprehensive experimental validation, including in the context of the problem addressed in this study.

\section{\label{sec:method} Methods and algorithms}

In this work, the problem of the reconstruction of wavefront distortions is addressed using the hybrid method (Section~\ref{sec:hybrid}) that combines deep learning with a classical iterative optimization algorithm. A neural network, whose architecture is detailed in Section~\ref{sec:net}, is trained on numerical data generated according to the procedure outlined in Section~\ref{sec:generator}. The trained neural network is then used to predict the wavefront distortions. Subsequently, the predicted distortions are decomposed into Zernike modes to serve as an initial guess for the L-BFGS optimization algorithm, which is described in Section~\ref{sec:lbfgs}. 

Although two beam cross-sections in the near-focal planes are theoretically sufficient to resolve the ambiguity of wavefront distortions, three measurement planes were applied in this study. Preliminary experiments have shown that correction of weak distortions using only two images leads to systematic errors, since this method tends to interpret noise, blur artifacts, and scaling inaccuracies as optical aberrations. To regularize the optimization, a third measurement with a defocus of the opposite sign was incorporated into the L-BFGS algorithm, significantly enhancing the robustness when processing experimental data. In all experiments, the neural network is trained on and processes only two images, whereas the L-BFGS algorithm performs the optimization based on all three images.

\subsection{\label{sec:generator} Data generation} 

\begin{figure}[H]
    \centering
    \includegraphics[width=0.5\textwidth]{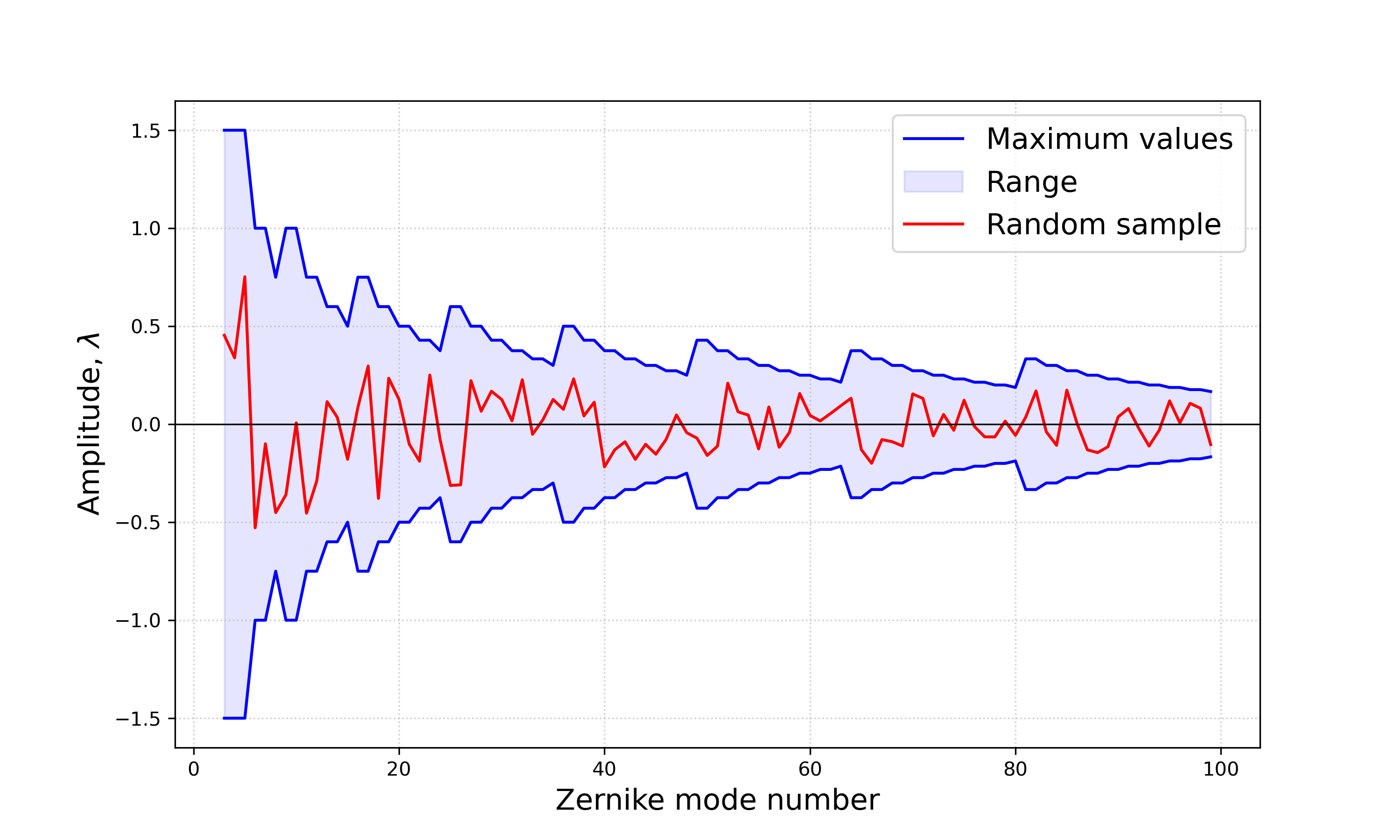}
    \caption{\label{fig:mode_range} Maximum and minimum bounds for the Zernike mode amplitudes, along with an example of a generated coefficient set.}
\end{figure}

The hybrid method takes as input the distributions of the electric field amplitude at the focal plane and at two symmetrically defocused planes, which are obtained by taking the square root of the intensity measured by the camera. The following procedure was used to generate the synthetic dataset. First, the near-field amplitude was defined by a uniform circular aperture. Next, the Zernike coefficients, denoted by $\bm{a}$, were randomly sampled from a uniform distribution $\mathcal{U}(-1, 1)$ and subsequently scaled by the inverse of their corresponding radial orders. The wavefront $W(\bm{a}, x, y)$ was constructed as a linear combination of these modes:

$$ W(\bm{a}, x, y) = \sum_{j=1}^{N} a_j Z_j(x,y), $$

\noindent where $Z_j(x,y)$ represents the $j$-th Zernike mode \cite{wyant1992basic}, $a_j$ is the corresponding Zernike coefficient, $x$ and $y$ are Cartesian coordinates, and $N$ is the total number of modes. The coefficients for the piston and tip/tilt terms were set to zero. The sampling bounds and a representative set of generated coefficients are illustrated in Fig.~\ref{fig:mode_range}.

To calculate the intensity distribution in the focal plane, the 2D Fast Fourier Transform was used to relate the complex field, including the amplitude distribution in the near field and the wavefront, to the intensity distribution in the focal plane. To simulate defocused intensity distributions, an additional phase term corresponding to the defocus Zernike mode was added to the wavefront with an amplitude of $\pm 0.5\lambda$. To enhance the sensitivity of the neural network to low-intensity features at the periphery of the focal spot, a square-root transformation was applied to the intensity distributions. Thus, for notational convenience, we define the inputs to the method as $\mathrm{FF}_j$:

$$ \mathrm{FF}_j(\bm{a}) = \left| \mathcal{F}  \left\{ A(x,y) \exp\left[ i 2\pi \left( W(\bm{a},x,y) + \alpha_j Z_3(x,y) \right) \right] \right\} \right|,$$

\noindent where $\mathcal{F}$ is the 2D Fourier transform operator, $A(x,y)$ is the aperture function, $W(\bm{a},x,y)$ is the wavefront expressed in units of wavelength, $Z_3(x,y)$ is the Zernike mode for defocus, and $\alpha_j$ is the corresponding defocus coefficient ($\alpha_1 = -0.5\lambda$, $\alpha_2 = 0$, $\alpha_3 = 0.5\lambda$).

\begin{figure}[H]
    \centering
    \includegraphics[width=0.44\textwidth]{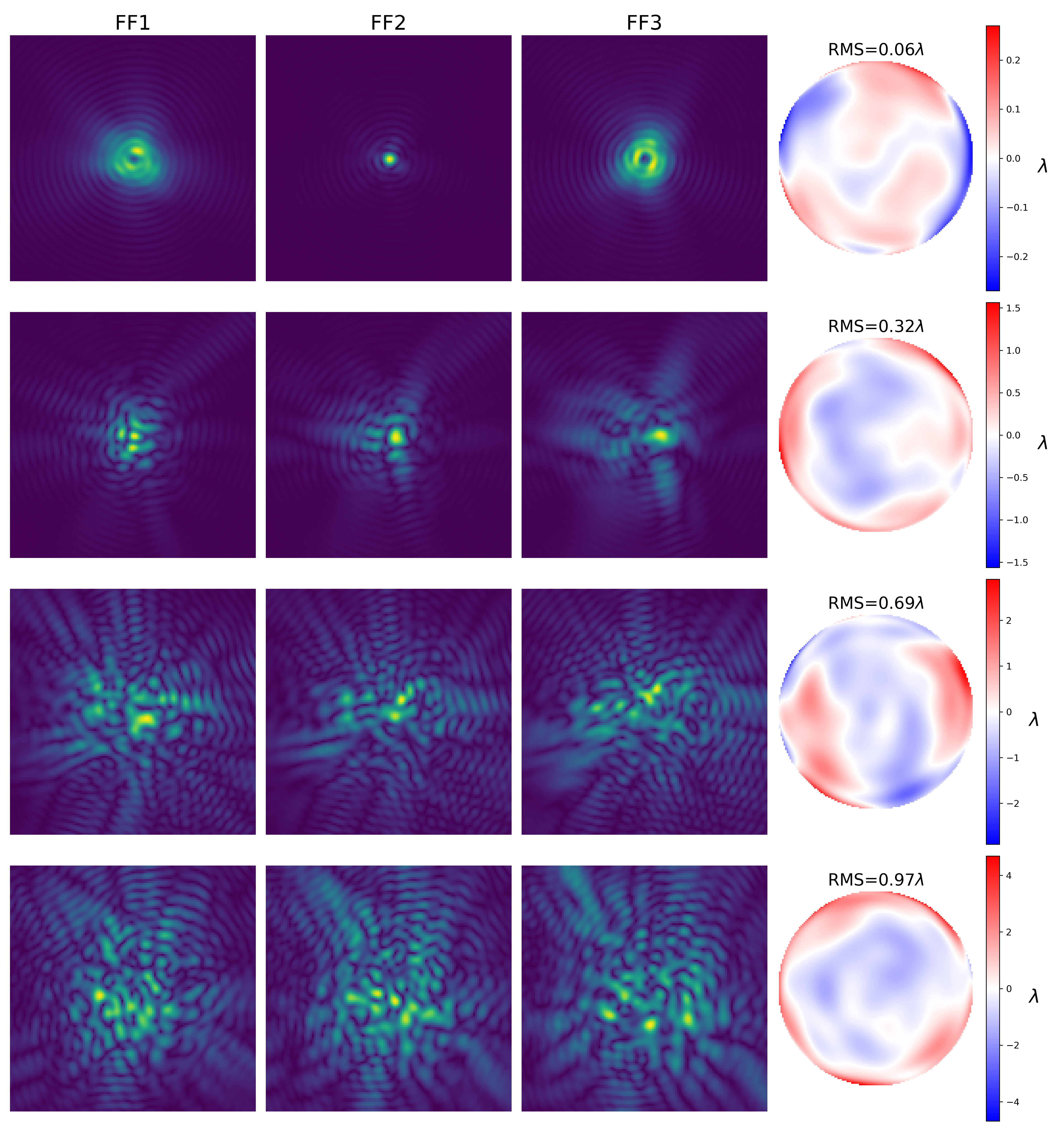}
    \caption{\label{fig:num_data} Examples of synthetic data. Columns (left to right): distributions of the electric field amplitude at the defocused plane ($\alpha_1 = -0.5\lambda$), at the focal plane ($\alpha_2 = 0\lambda$), and at the defocused plane ($\alpha_3 = 0.5\lambda$); corresponding wavefront distortions.}
\end{figure}

Following generation, the focal and defocused images were center-cropped to a spatial resolution of $128 \times 128$ pixels. The effective pixel size was defined as $d_{\text{pix}} = 0.25\lambda F/D$, where $F$ is the focal length and $D$ is the beam diameter. The total field of view for each image corresponded to $32\lambda F/D \times 32\lambda F/D$. Examples of the synthetic data generated for various aberration strengths using 100 Zernike modes are presented in Fig.~\ref{fig:num_data}. The final dataset consisted of training set of 100,000 samples, validation set of 5,000 samples and test set of 1,000 samples.

\subsection{\label{sec:net} Deep learning model architecture}

\begin{figure}[h!]
    \centering
    \includegraphics[width=0.48\textwidth]{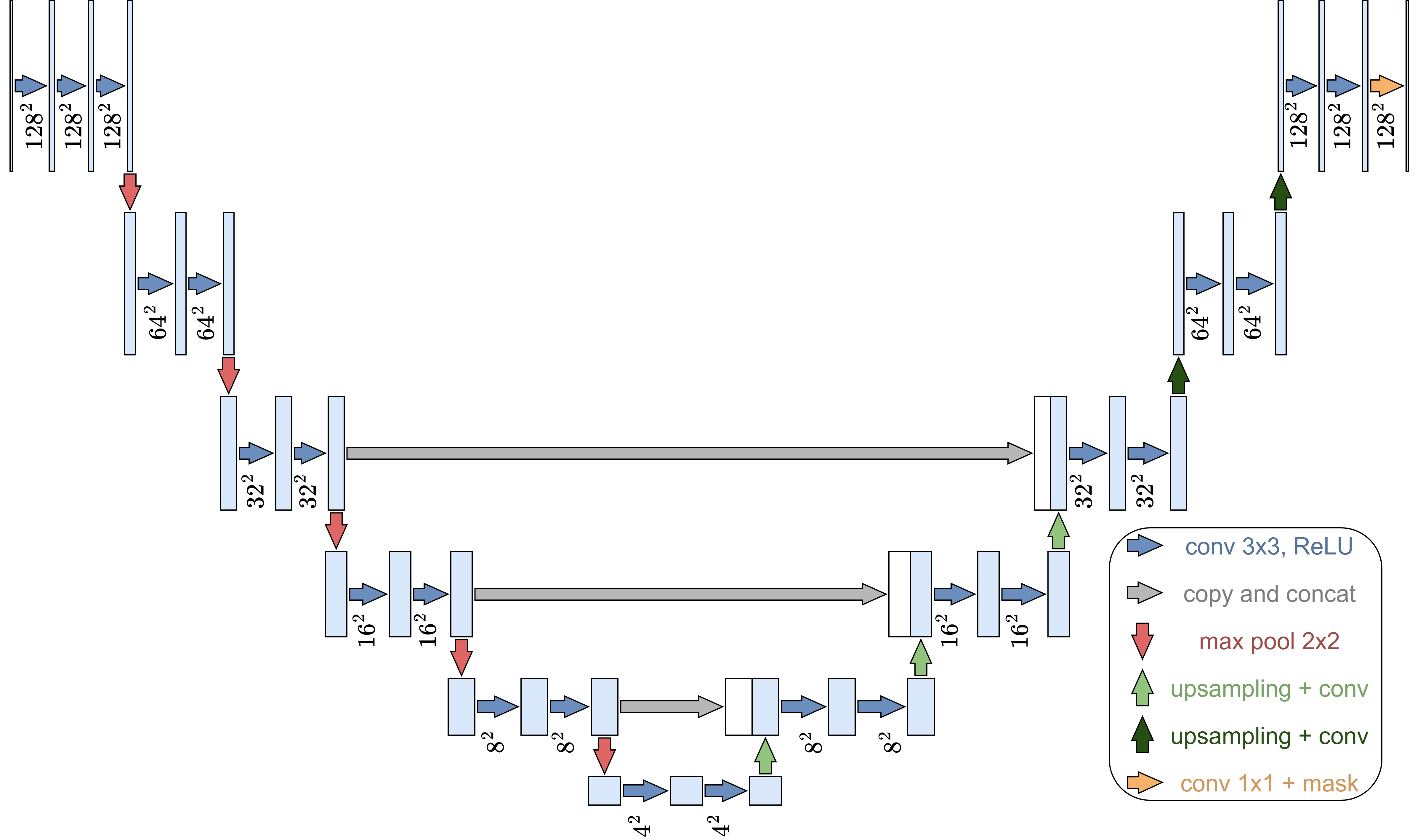}
    \caption{\label{fig:Unet} Architecture of the modified U-Net applied for the reconstruction of wavefront distortions.}
\end{figure}

A neural network $\hat{W} = NN(\mathrm{FF}_1, \mathrm{FF}_2)$ is applied to reconstruct the wavefront distortions $\hat{W}$ from two distributions of the electric field amplitude, one at the defocused plane $\mathrm{FF}_1$ and one at the focal plane $\mathrm{FF}_2$. A modified U-Net architecture \cite{ronneberger2015u}, illustrated in Fig.~\ref{fig:Unet}, was used for this task. The original U-Net was designed for semantic segmentation, where the input and output domains share spatial correspondence, unlike our problem. Consequently, the early-stage skip connections were removed.

The distributions of the electric field amplitude at the focal and defocused planes were concatenated along the channel dimension to form a two-channel input tensor. The network's output was subsequently multiplied by a circular mask. The model was trained to predict the pixel-wise wavefront distortions directly. The network architecture comprises five sequential encoder blocks, a bottleneck layer, and five decoder blocks. The encoder and decoder pathways are bridged by skip connections, which are retained only for the last three encoder blocks, passing their feature maps to the corresponding decoder stages.

Each encoder block consists of two $3 \times 3$ convolutional layers with a padding of 1, each followed by a non-linear activation function, and terminates with a $2 \times 2$ max pooling layer for spatial downsampling.

The first three decoder blocks use transposed convolutions for spatial upsampling, followed by concatenation with the feature maps routed from the corresponding encoder. The combined feature maps are then processed by a sequence of two $3 \times 3$ convolutions and activations, mirroring the structure of the encoder blocks. The final two decoder blocks, lacking skip connections, rely solely on the output of the preceding decoder stage. These final blocks use bilinear upsampling followed by a convolutional block identical to those in the encoder.

The model was optimized using a mean squared error loss function. Training was performed for 600 epochs with a batch size of 500 using the Adam optimizer \cite{kingma2014adam} at a learning rate of $5 \times 10^{-4}$. No regularization was applied. The ReLU activation function was used across all hidden layers, while the final output layer used a linear activation. The model was trained on the synthetic dataset generated using 100 Zernike modes.

\subsubsection{\label{sec:aug} Data preprocessing and augmentation}

Each distribution of the electric field amplitude $\mathrm{FF}_i$ was independently normalized by its maximum value. The target wavefront distortions were normalized by a factor of $3\lambda$, yielding a value range of approximately $[-3, 3]$, which is conducive to stable training of the neural network. During inference, the inverse scaling was applied to the predicted wavefront distortions.

Data augmentation was employed to prevent overfitting to the idealized synthetic data. Four types of transformations were used to simulate measurement errors: spatial translation, Gaussian blurring, quantization, and additive uniform noise. Each transformation was applied independently with a probability of 50\%. For each training batch, a random scaling coefficient $r \sim \mathcal{U}(0, 1)$ was generated to modulate the perturbation amplitude. This exposed the model to data with varying error magnitudes, ranging from low levels, which accelerate convergence in the early training stages, to levels approaching real experimental conditions.

The distributions were randomly shifted by a value sampled from $\mathcal{U}(-7r, 7r)$ pixels, where the effective pixel size is $d_{\text{pix}} = 0.25\lambda F/D$. This addition allows the model to adapt to beam displacements relative to the image. Blurring was performed using a $5 \times 5$ Gaussian kernel with a standard deviation sampled from $\mathcal{U}(0.01r, 0.5r)$. To emulate the limited dynamic range and analog-to-digital discretizations of the camera, the electric field amplitude quantization was modeled by converting the data from 32-bit floating point to 8-bit floating point and back. Finally, additive uniform noise sampled from $\mathcal{U}(0, 0.07r)$ was superimposed on the electric field amplitude.

\subsubsection{Circular buffer}

The synthetic dataset used for model training was not static. The dataset was dynamically updated during the optimization process. The data were stored in a fixed-size circular buffer, where older examples were continuously replaced with newly generated ones \cite{rodimkov2026using}. The training procedure followed a standard pipeline. Mini-batches were sampled from the buffer to compute the forward and backward passes for the weight updates.

In parallel with model training, new synthetic data were continuously generated via numerical simulations and appended to the circular buffer. The generated data were written to the buffer in chunks of 100 samples. Insertion and extraction operations on the buffer were strictly synchronized to ensure that only one operation was executed at a given time.

The buffer capacity was set to 100,000 samples and was completely pre-filled before the start of model training. The data generation pipeline was parallelized across 10 processes. On average, approximately 15,000 samples in the buffer were updated per training epoch.

\subsection{\label{sec:lbfgs} L-BFGS Algorithm}

The L-BFGS algorithm \cite{liu1989limited} is a quasi-Newton method for unconstrained optimization, designed to find a local optimum of a multivariate function. As a local optimization algorithm, its outcome depends on the initial guess, and it does not guarantee convergence to a global optimum. Unlike the exact Newton method, the L-BFGS does not require explicit storage or inversion of the Hessian.

In the standard Newton method, an iterative procedure determines the search direction using the exact inverse Hessian, $H_k = (\nabla^2 f(\theta_k))^{-1}$, yielding the update rule $\theta_{k+1} = \theta_k - \alpha_k H_k g_k$, where $g_k = \nabla f(\theta_k)$ is the gradient. The defining feature of the L-BFGS is that it approximates the inverse Hessian $H_k$ using only the $m$ most recent pairs of parameter differences $s_i = \theta_{i+1} - \theta_i$ and gradient differences $d_i = g_{i+1} - g_i$. The search direction $p_k = -H_k g_k$ is computed using the Two-loop Recursion algorithm (Algorithm~\ref{alg:lbfgs}) \cite{nocedal2006numerical}.

\begin{algorithm}[H]
\caption{L-BFGS Two-loop Recursion}\label{alg:lbfgs}
\begin{algorithmic}[1]
\State \textbf{Input:} Current gradient $g_k$, history of $m$ pairs $\{s_i, d_i\}_{i=k-m}^{k-1}$
\State \textbf{Output:} Descent direction $p_k$
\State $q \gets g_k$
\For{$i = k-1$ \textbf{down to} $k-m$}
    \State $\rho_i \gets \frac{1}{d_i^T s_i}$
    \State $\alpha_i \gets \rho_i s_i^T q$
    \State $q \gets q - \alpha_i d_i$
\EndFor
\State $H^0_k \gets \frac{s_{k-1}^T d_{k-1}}{d_{k-1}^T d_{k-1}} I$
\State $z \gets \gamma_k q$
\For{$i = k-m$ \textbf{to} $k-1$}
    \State $\beta \gets \rho_i d_i^T z$
    \State $z \gets z + s_i (\alpha_i - \beta)$
\EndFor
\State \Return $p_k \gets -z$ \Comment{$p_k$ approximates $-H_k g_k$}
\end{algorithmic}
\end{algorithm}

The parameters are then updated as $\theta_{k+1} = \theta_k + \alpha_k p_k$. In this work, we use the L-BFGS implementation from the PyTorch framework. The stopping criteria are set to a gradient norm tolerance of $10^{-7}$ and a relative change in the objective function of $10^{-9}$. The step size $\alpha_k$ is determined adaptively via a line search satisfying the Wolfe conditions.

This algorithm is used to optimize the amplitudes of the Zernike modes, the number of which varies depending on the experiment. The optimization objective is the similarity between the ground-truth image and the image numerically simulated from the predicted Zernike coefficients. To evaluate this, the maximum of the normalized cross-correlation ($NCC_{\text{max}}$) is used:

\begin{eqnarray}
&& NCC_{\text{max}}(\mathrm{FF}^{\text{pred}}, \mathrm{FF}^{\text{truth}}) = \frac{1}{n} \max \nonumber\\ 
&&\left( \text{Re} \left[ \mathcal{F}^{-1} \left( \mathcal{F}(\tilde{\mathrm{FF}}^{\text{pred}}) \cdot \mathcal{F}(\tilde{\mathrm{FF}}^{\text{truth}})^* \right) \right] \right) \nonumber
\end{eqnarray}

\noindent where $\tilde{\mathrm{FF}}^{\text{pred}} = \frac{\mathrm{FF}^{\text{pred}} - \mu_{\text{pred}}}{\sigma_{\text{pred}}}$ and $\tilde{\mathrm{FF}}^{\text{truth}} = \frac{\mathrm{FF}^{\text{truth}} - \mu_{\text{truth}}}{\sigma_{\text{truth}}}$ are the normalized images; $\mathrm{FF}^{\text{pred}}$ is the simulated image generated from the predicted amplitudes; $\mathrm{FF}^{\text{truth}}$ is the ground-truth image; $\mu_{\text{pred}}$, $\mu_{\text{truth}}$, $\sigma_{\text{pred}}$, and $\sigma_{\text{truth}}$ denote the respective means and standard deviations of $\mathrm{FF}^{\text{pred}}$ and $\mathrm{FF}^{\text{truth}}$; $\mathcal{F}$ and $\mathcal{F}^{-1}$ represent the 2D forward and inverse Fourier transforms, respectively; $^*$ denotes complex conjugation; $\text{Re}[\cdot]$ extracts the real part of a complex number; and $n$ is the total number of pixels in the image. This metric is translation-invariant, which is crucial in experimental where precise centering of the data is unfeasible.

To optimize the Zernike mode amplitudes $\bm{a}$, the negative sum of the normalized cross-correlations across three images is minimized: 

$$
f(\bm{a}) = -\sum_{j=1}^3 NCC_{\text{max}}(\mathrm{FF}_j^{\text{pred}}(\bm{a}), \mathrm{FF}_j^{\text{truth}}).
$$

\subsection{\label{sec:hybrid} Hybrid Method}

\begin{figure*}
    \centering
    \includegraphics[width=0.95\textwidth]{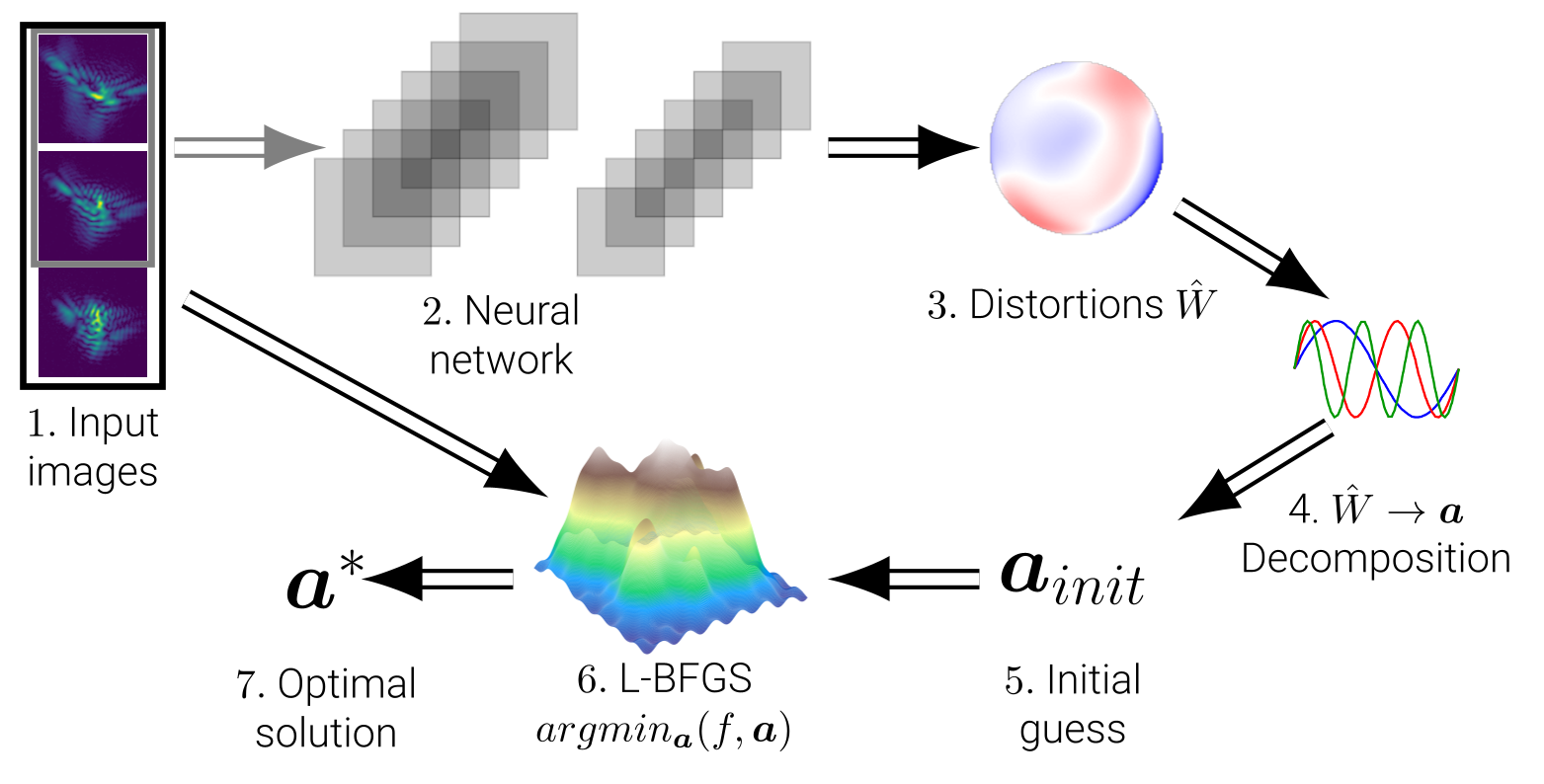}
    \caption{\label{fig:scheme_hybrid} Schematic of the workflow of the hybrid method.}
\end{figure*}

The hybrid method is used to predict the Zernike mode coefficients, as illustrated in Fig.~\ref{fig:scheme_hybrid}.In the first stage, the trained convolutional neural network reconstructs per-pixel wavefront distortions based on intensity distributions obtained in the focal and defocused planes, denoted as $\hat{W} = NN(\mathrm{FF}_1, \mathrm{FF}_2)$. The reconstructed wavefront distortions $\hat{W}$ are then decomposed into a selected number of Zernike coefficients $\bm{a}_{\text{init}}$. These initial estimates $\bm{a}_{\text{init}}$ serve as the starting point for the L-BFGS algorithm, which minimizes the objective function $f(\bm{a})$, defined as the negative sum of the normalized cross-correlations between the numerically simulated distributions of the electric field amplitude and the corresponding experimentally measured intensity distributions.

A key advantage of this hybrid method is that the neural network and the L-BFGS algorithm do not require identical input data. This decoupling means the neural network can be trained once using a standard two-plane phase diversity scheme to provide a robust initial guess. The subsequent optimization can then incorporate an arbitrary set of measurement planes to stabilize the solution and achieve the desired reconstruction accuracy. Consequently, this method offers significant flexibility. It naturally accommodates additional observation planes or alternative phase diversity configurations specific to a given experimental setup, entirely circumventing the computationally expensive process of retraining the neural network.

\section{\label{sec:result} Results}

\subsection{Numerical results}

First, we compared the accuracy and computational time of four approaches for the reconstruction of wavefront distortions using numerical data. The first approach employs a trained convolutional neural network (NN) to predict the wavefront distortions. The second approach uses the L-BFGS algorithm to optimize the Zernike coefficients with zero initialization (L-BFGS (zero init)). This initial guess is physically justified, as a wavefront without aberrations corresponds to a vector of zero coefficients. The third approach relies on the L-BFGS algorithm executed with 10 random initializations (L-BFGS ($10 \times \text{random}$)). The procedure for generating these starting points is identical to the data generation procedure used in the data generation. The fourth approach is a hybrid method, which employs a convolutional neural network for the initial reconstruction of wavefront distortions. The resulting distortions are then projected onto the Zernike basis to provide an initial guess for the L-BFGS algorithm (Hybrid method). To ensure consistency, all methods reconstructed the amplitudes of Zernike modes, i.e., the wavefront distortions were decomposed into Zernike modes when necessary. This step is essential because the primary objective of this work is the calibration of the AOS, where additional wavefront corrections are applied via Zernike modes.

Experiments were conducted across three configurations. Test data were generated using 25, 50, and 100 Zernike modes, and all approaches reconstructed the corresponding number of coefficients in each trial. The RMS of the residual wavefront distortions was used to evaluate the reconstruction accuracy. To calculate this, the predicted Zernike coefficients were converted into wavefront distortions, and the RMS of the residual difference between the predicted and ground-truth wavefront distortions was computed. The error metrics and computation times for all approaches are summarized in Table~\ref{tab:result_num}. Notably, the hybrid method achieves the highest accuracy, particularly as the number of Zernike modes increases. The neural network also demonstrates acceptable performance, although its RMS error is approximately five times higher than that of the hybrid method.The L-BFGS algorithm with zero or random initialization shows significantly worse performance. Their RMS errors are an order of magnitude higher than those of the hybrid method. This substantial performance gap underscores the critical importance of an initial guess.

\begin{table*}
\caption{\label{tab:result_num} Accuracy and computation time of approaches for different numbers of Zernike coefficients.}
\centering
\begin{tabular}{|l|cc|cc|cc|}
\hline
& \multicolumn{2}{c|}{25 Zernike mods} & \multicolumn{2}{c|}{50 Zernike mods} & \multicolumn{2}{c|}{100 Zernike mods} \\
\cline{2-7}
Method & RMS ($\lambda$) & Time (s) & RMS ($\lambda$) & Time (s) & RMS ($\lambda$) & Time (s) \\
\hline
Net         & $6.05 \cdot 10^{-2}$ & 0.05 & $6.68 \cdot 10^{-2}$ & 0.05 & $1.09 \cdot 10^{-1}$ & 0.05 \\
L-BFGS (zero init)  & $2.45 \cdot 10^{-1}$ & 3.08 & $3.59 \cdot 10^{-1}$ & 3.68 & $4.91 \cdot 10^{-1}$ & 6.61 \\
L-BFGS ($10\times \text{random}$) & $7.82 \cdot 10^{-2}$ & 38.43 & $3.10 \cdot 10^{-1}$ & 65.25 & $5.35 \cdot 10^{-1}$ & 105.16 \\
Hybrid method   & $1.33 \cdot 10^{-2}$ & 1.17 & $9.56 \cdot 10^{-3}$ & 1.36 & $3.16 \cdot 10^{-2}$ & 2.70 \\
\hline
\end{tabular}
\end{table*}

Computation time measurements performed on a CPU (Intel Core i5-10400) indicate that the neural network achieves the fastest inference. A single run of the L-BFGS algorithm takes an average of 3-6 seconds, depending on the number of Zernike coefficients being optimized. As the dimensionality of the optimization parameter space increases, the problem complexity grows, leading to an increased number of iterations. The hybrid method reduces the overall computation time by 2-3 times, since the neural network provides an initial guess close to the local optimum. Consequently, the number of iterations required for the convergence of the L-BFGS algorithm decreases. Utilizing GPU acceleration can significantly reduce the computation time for all approaches, yielding a remarkable speedup of over 4 times faster.

\begin{figure}[H]
    \centering
    \begin{subfigure}{0.45\textwidth}
        \centering
        \includegraphics[width=\textwidth]{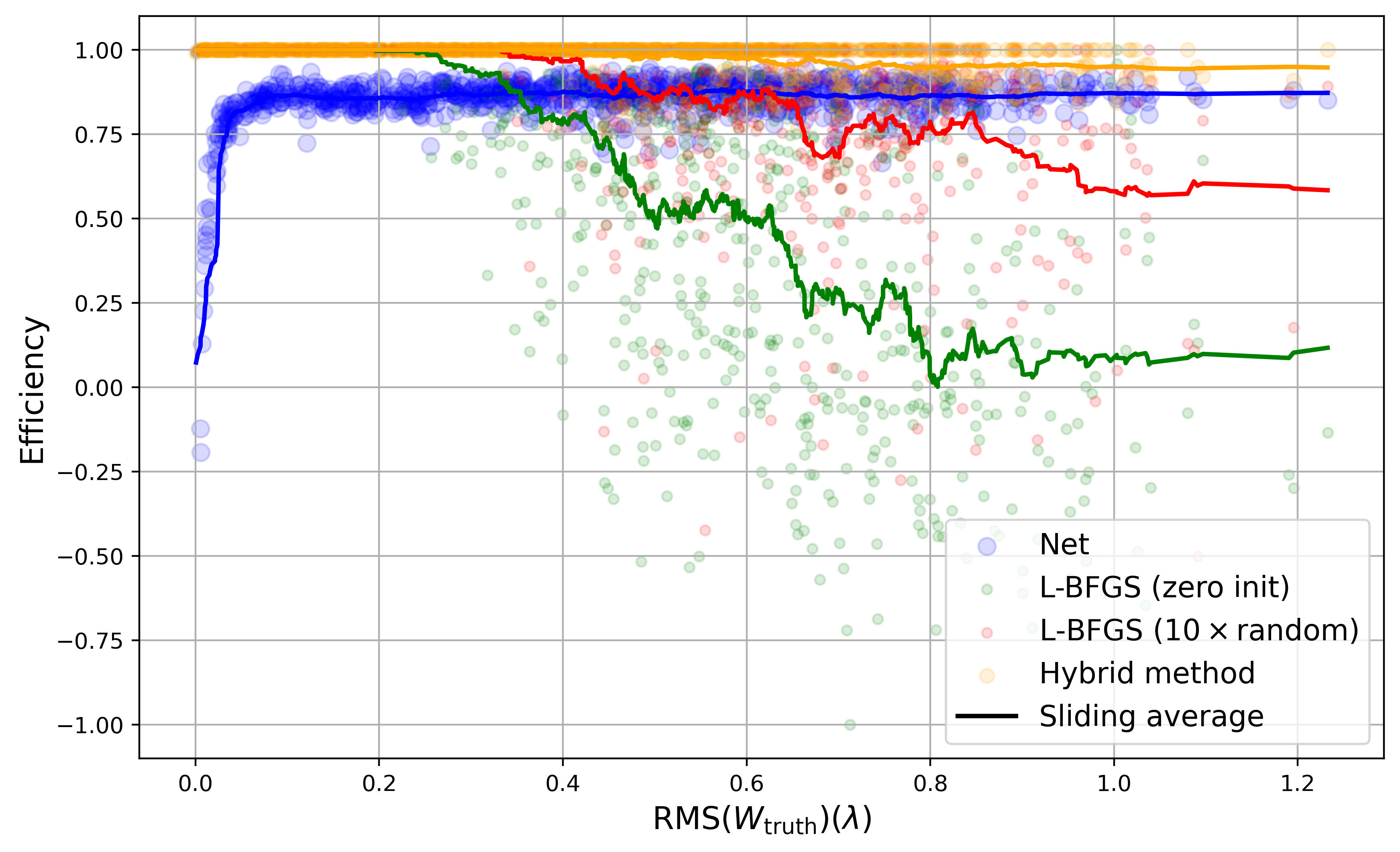}
        \subcaption{}
    \end{subfigure}
    \vspace{0.5cm}
    \begin{subfigure}{0.45\textwidth}
        \centering
        \includegraphics[width=\textwidth]{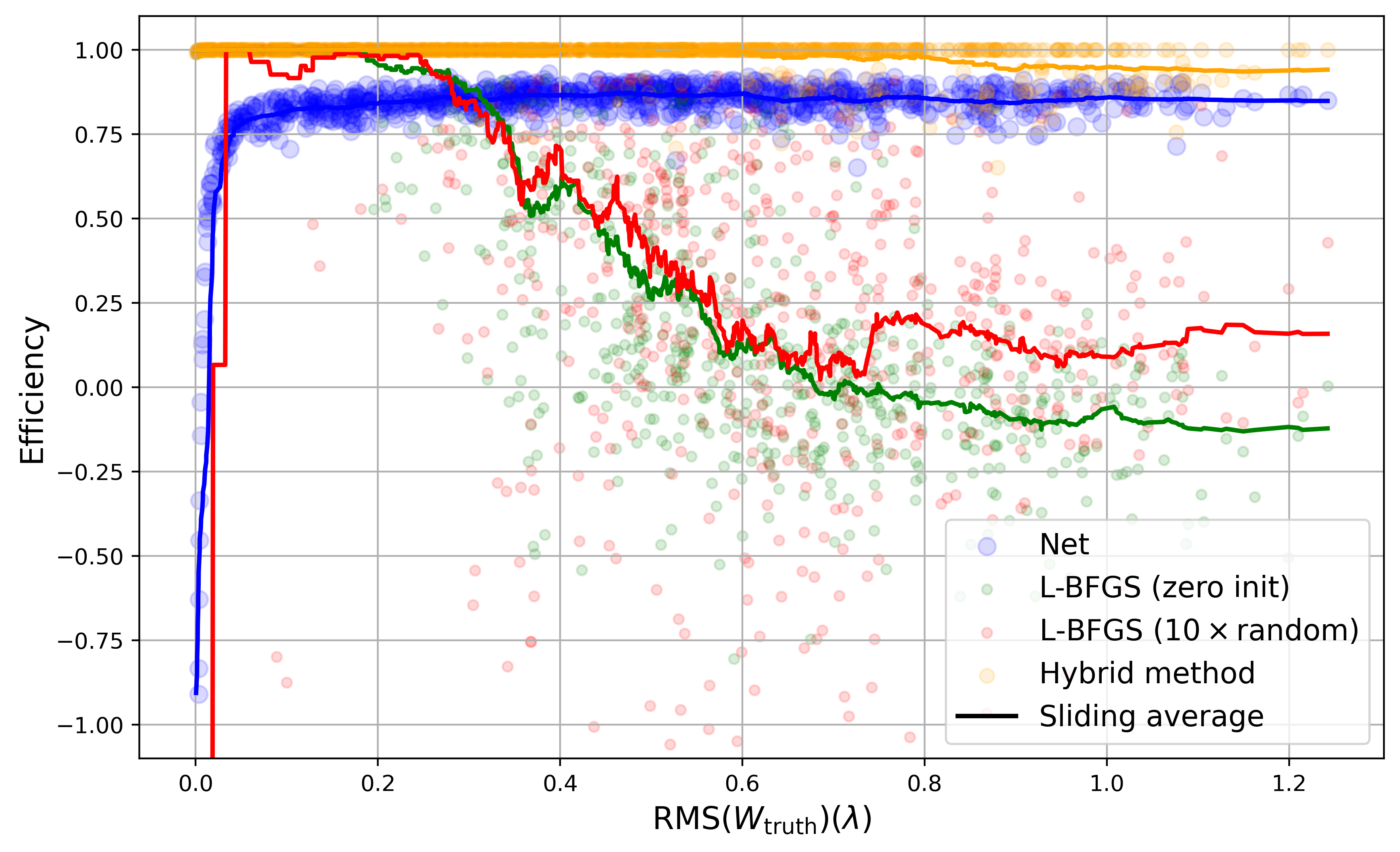}
        \subcaption{}
    \end{subfigure}
    \vspace{0.5cm}
    \begin{subfigure}{0.45\textwidth}
        \centering
        \includegraphics[width=\textwidth]{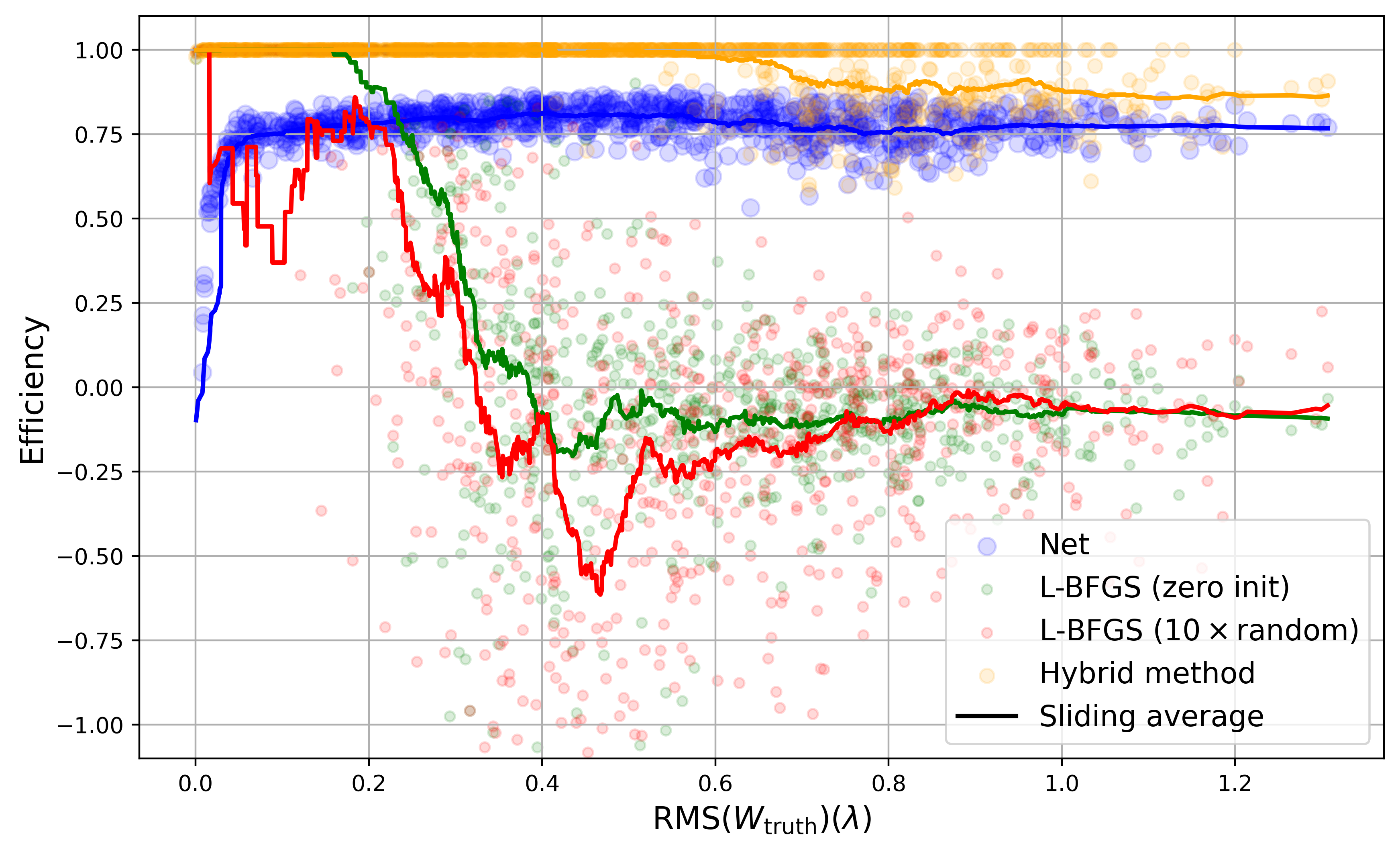}
        \subcaption{}
    \end{subfigure}
    \caption{\label{fig:num_result} Dependence of algorithm efficiency on distortion level for different numbers of Zernike modes used for data generation. (a) 25 modes, (b) 50 modes, (c) 100 modes. The colored line corresponds to a moving average with a window size of 50 points.}
\end{figure}

Figure~\ref{fig:num_result} illustrates the dependence of the reconstruction efficiency on the magnitude of the wavefront distortions for different numbers of Zernike modes used in the data generation. The efficiency is defined as:

\begin{equation*}
    \text{efficiency} = 1 - \frac{\text{RMS}(\hat{W}_{\text{pred}} - W_{\text{truth}})}{\text{RMS}(W_{\text{truth}})},
\end{equation*}

\noindent where $\hat{W}_{\text{pred}}$ denotes the predicted wavefront distortions and $W_{\text{truth}}$ represents the ground-truth wavefront distortions.

The neural network demonstrates high robustness, maintaining high efficiency over a wide range of RMS wavefront distortion values, except for cases with exceedingly small initial wavefront distortions. This behavior occurs because the relative metric becomes highly sensitive to small absolute errors at low RMS levels. Furthermore, the reconstruction accuracy improves as the number of modes decreases, which is attributed to the reduced spatial complexity of the target wavefront distortions. The L-BFGS algorithm exhibits high, near-ideal efficiency in the area of small wavefront distortions for both zero and random initializations. However, as the amplitude of the wavefront distortions or the number of modes increases, the frequency of cases yielding low or negative efficiency grows. In certain instances, the efficiency takes large negative values. This effect is observed, for example, in the L-BFGS algorithm with random initialization in the regime of small wavefront distortions (Fig.~\ref{fig:num_result}(b)). The average efficiency of the L-BFGS algorithm for both initialization strategies begins to degrade substantially at an RMS of wavefront distortions exceeding $0.35\lambda$. The hybrid method consistently delivers high accuracy across almost all evaluated scenarios, regardless of the number of modes or the RMS of the initial wavefront distortions. The efficiency of the hybrid method exceeds $0.99$ in 80\% of the cases. Moreover, fewer than 2\% of the samples exhibit an efficiency below $0.9$ for wavefront distortions with an RMS below $0.6\lambda$, which constitutes the target operational range for the physical experiment. 

The hybrid method demonstrated the best accuracy and stability while maintaining sufficiently fast processing times. In the next part of the study, the method was subsequently tested for AOS calibration under real-world experimental conditions.

\subsection{Experiment}

\subsubsection{Experimental facility}

\begin{figure}[h]
\centering
\includegraphics[width=0.48\textwidth]{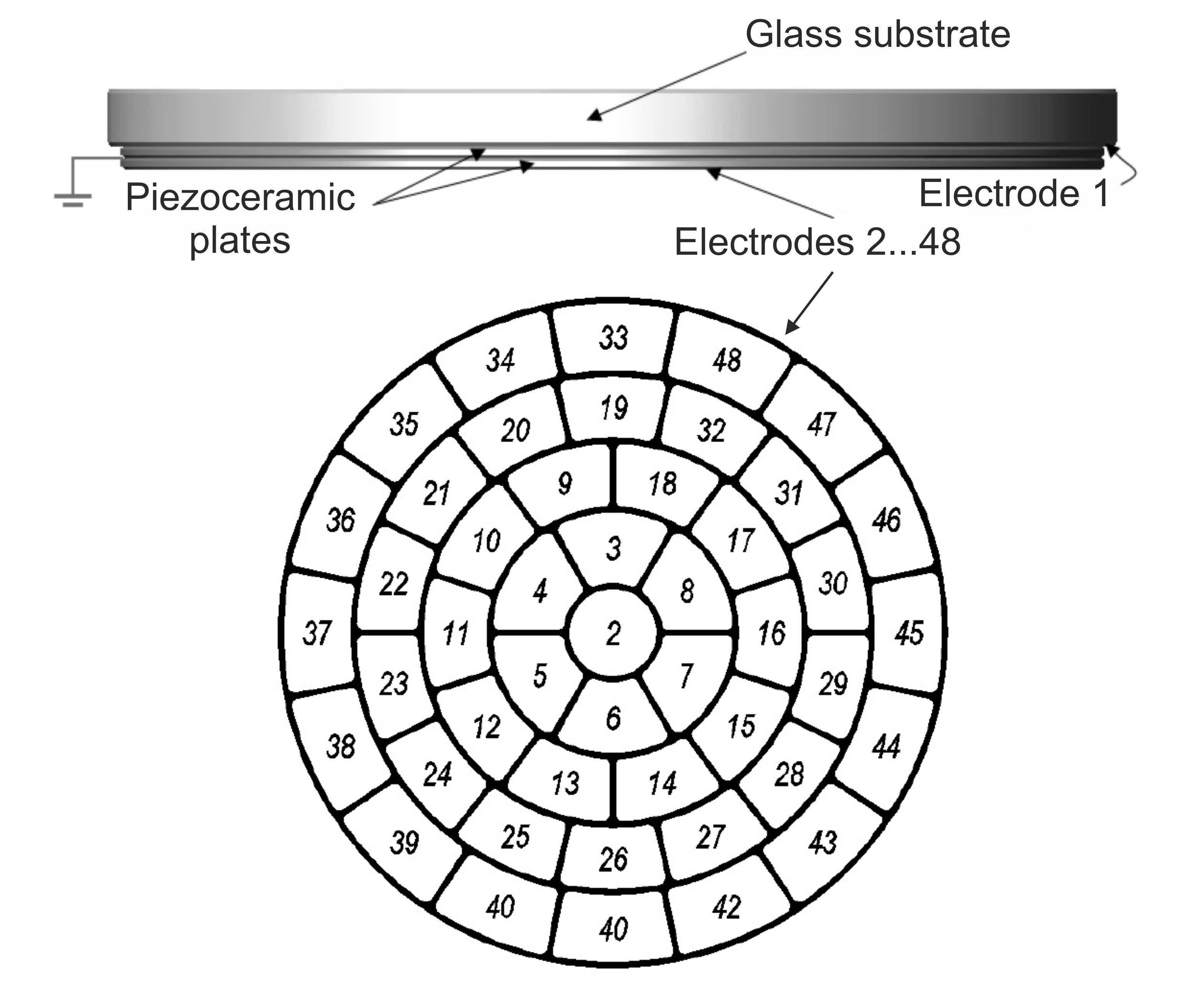}
\caption{\label{fig:DM} Schematic of the deformable mirror and the electrode layout on the outer piezoceramic plate.}
\end{figure}

The experiments were performed using the AOS \cite{rukosuev2002adaptive, soloviev2020adaptive, akaoptics} consisting a of Shack-Hartmann WFS and a bimorph DM. The DM design and electrode geometry are illustrated in Fig.~\ref{fig:DM}. The DM is a three-layer composite structure consisting of a polished substrate with a reflective coating and two piezoceramic disks. All mirror components are rigidly bonded.

The operation of the DM relies on the inverse piezoelectric effect: applying a control voltage to the piezoceramic disks induces longitudinal compression or expansion of the piezoceramic material. This, in turn, deforms the DM surface, enabling high-precision shape control. The inner piezoelectric disk with a solid electrode is used to form the overall curvature of the DM surface. The outer surface of the second piezoceramic disk is patterned with galvanically isolated segmented electrodes to generate localized surface deformations. In this study, a 48-electrode DM with a 120~mm diameter was used, designed for a 100~mm aperture. The DM substrate was coated with a multilayer dielectric coating, ensuring a reflectivity of at least $99.9\%$ in the $910 \pm 40$~nm spectral range.

The Shack-Hartmann WFS consisted of a CMOS camera coupled with a microlens array. The camera sensor had a clear aperture of approximately 1.1~cm and a pixel size of 5.5~\textmu m. The microlens array had a focal length of 3.2~mm and a lenslet size of 136~\textmu m. The WFS provided a residual wavefront measurement error with an RMS below $0.025\lambda$. The operating principle of the Shack-Hartmann WFS involves spatially sampling the incident beam into subapertures using the microlens array. Each lenslet focuses the light onto the camera sensor, forming a two-dimensional array of focal spots known as a Hartmannogram. The displacements of the focal spot centroids relative to their reference positions allow for the computation of local wavefront slopes, from which the wavefront distortions are reconstructed. The measured wavefront distortions represent the difference between the incident wavefront and a calibrated reference wavefront. When measuring wavefront distortions in beams with an aperture larger than the WFS sensor, an telescopic system is used. In such configurations, a proper closed-loop AOS operation requires precise optical conjugation between the DM and the WFS planes.

The standard wavefront correction procedure in adaptive optics is implemented via phase conjugation in a closed-loop control system. This process operates as an iterative cycle:
\begin{enumerate}
\item The WFS measures the wavefront distortions, defined as the wavefront difference between the incident beam and the reference wavefront.
\item The control system calculates the corresponding command voltages based on the WFS measurements and applies them to the DM.
\item The DM updates its surface profile in response to the applied control signals.
\item This cycle repeats continuously to ensure a dynamic wavefront stabilization.
\end{enumerate}
Through this feedback loop, the AOS minimizes the residual wavefront distortions between the incident and reference wavefronts.

\begin{figure*}
\centering
\includegraphics[width=0.75\textwidth]{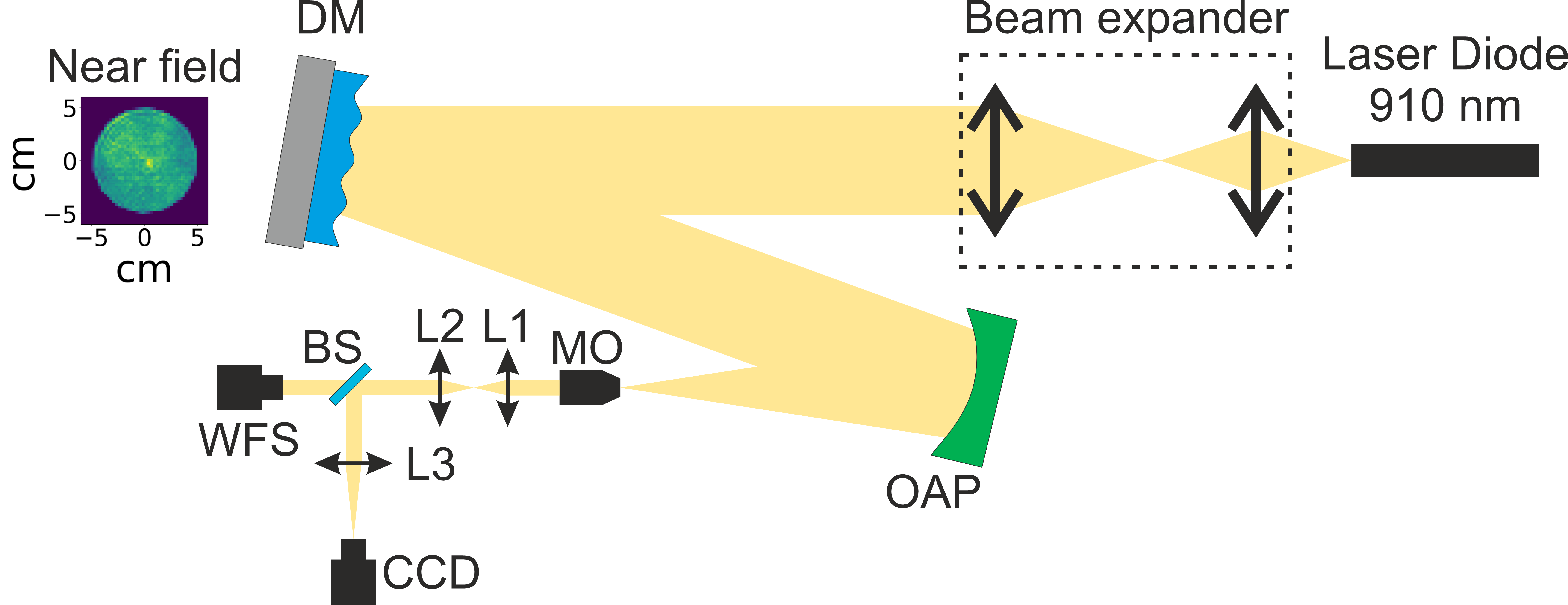}
\caption{\label{fig:exp_scheme} Schematic of the experimental setup. DM: deformable mirror; OAP: off-axis parabolic mirror; MO: microscope objective; BS: beam splitters; L1-L3: lenses; WFS: wavefront sensor; CCD: CCD camera.}
\end{figure*}

The experiment was performed at the PEARL laser facility \cite{soloviev2024research}. The optical scheme is depicted in Fig.~\ref{fig:exp_scheme}. The output from a laser diode was expanded by a beam expander to a clear aperture of approximately 10~cm. The beam was then reflected by the DM and focused by an off-axis parabolic (OAP) mirror with a focal length of $F = 760$~mm. The converging beam was collected by a microscope objective (MO) and subsequently collimated and demagnified to a 5~mm diameter using a relay lens system (L1-L2). A beam splitter (BS) divided the beam: the transmitted portion was directed into the WFS, while the reflected portion was focused by lens L3 onto a CCD camera for focal spot monitoring.

Phase diversity was implemented directly via the DM. This approach offers several significant advantages:
\begin{enumerate}
\item Utilizing the deformable mirror allows for a highly flexible phase diversity scheme. For instance, diverse types and amplitudes of wavefront distortions can be applied, enabling the acquisition of intensity distributions at an arbitrary number of focal and defocused planes. The reconstruction accuracy is then primarily limited by the spatial fidelity of the mirror's surface shaping capabilities.

\item No supplementary optical components are required in the measurement scheme, avoiding parasitic wavefront distortions and simplifying the optical alignment. Conventional methods, such as introducing an auxiliary focusing channel or a beam splitter, inherently induce inter-channel wavefront distortions. Alternatively, translating the camera along the optical axis demands high-precision motorized stages and degrades the temporal resolution of the system.

\item This approach seamlessly integrates into the existing AOS architecture, as it can be executed purely as a software procedure within the closed-loop control system.
\end{enumerate}

In each experiment, we introduced controlled wavefront distortions into the beam by adding combinations of Zernike modes to the wavefront. Zernike mode coefficients were generated similar to the synthetic data. Consequently, the target reference wavefront was shifted to a range with an RMS of wavefront distortions up to $0.6\lambda$, which corresponds to the limits of the DM. Following the introduction of these distortions, the AOS engaged in a closed-loop correction to minimize the residual wavefront distortions between the measured and the modified reference wavefronts. Once the control loop converged, the intensity distributions at the focal and defocused planes were recorded.

The experimentally acquired intensity distributions were subjected to a preprocessing pipeline. To suppress background noise, the median pixel value of a peripheral background region was subtracted from each distribution, and any resulting negative values were clipped to zero. Each distribution was then spatially centered based on its center of mass and normalized by its maximum value. Following this, an element-wise square root operation was applied. These preprocessed distributions of the electric field amplitude were subsequently fed into the hybrid method to reconstruct the corresponding wavefront distortions. The number of Zernike modes used was limited by the AOS software and set to 25.

\subsubsection{Experimental results}

The performance of the hybrid method was evaluated in nine experimental series, each with different initial wavefront distortions. Each series comprised nine correction iterations based on the predictions of the hybrid method. In applications aimed at maximizing the peak intensity, the most standard metric for assessing the compensation quality of wavefront distortions is the Strehl ratio. It is the ratio of the on-axis energy density measured at the focal waist to the maximum value achievable for the given near-field distribution, assuming a perfectly flat wavefront and ideal focusing optics. Thus, the deviation of the Strehl ratio from unity quantifies the degradation in the peak intensity at the focal plane caused by residual wavefront distortions. In addition to the Strehl ratio, the experiment also analyzed the residual wavefront distortions, calculated relative to the wavefront corresponding to the best focusing with a Strehl ratio of 0.98 (4th series, 5th iteration).

\begin{figure*}
    \centering
    \includegraphics[width=0.85\textwidth]{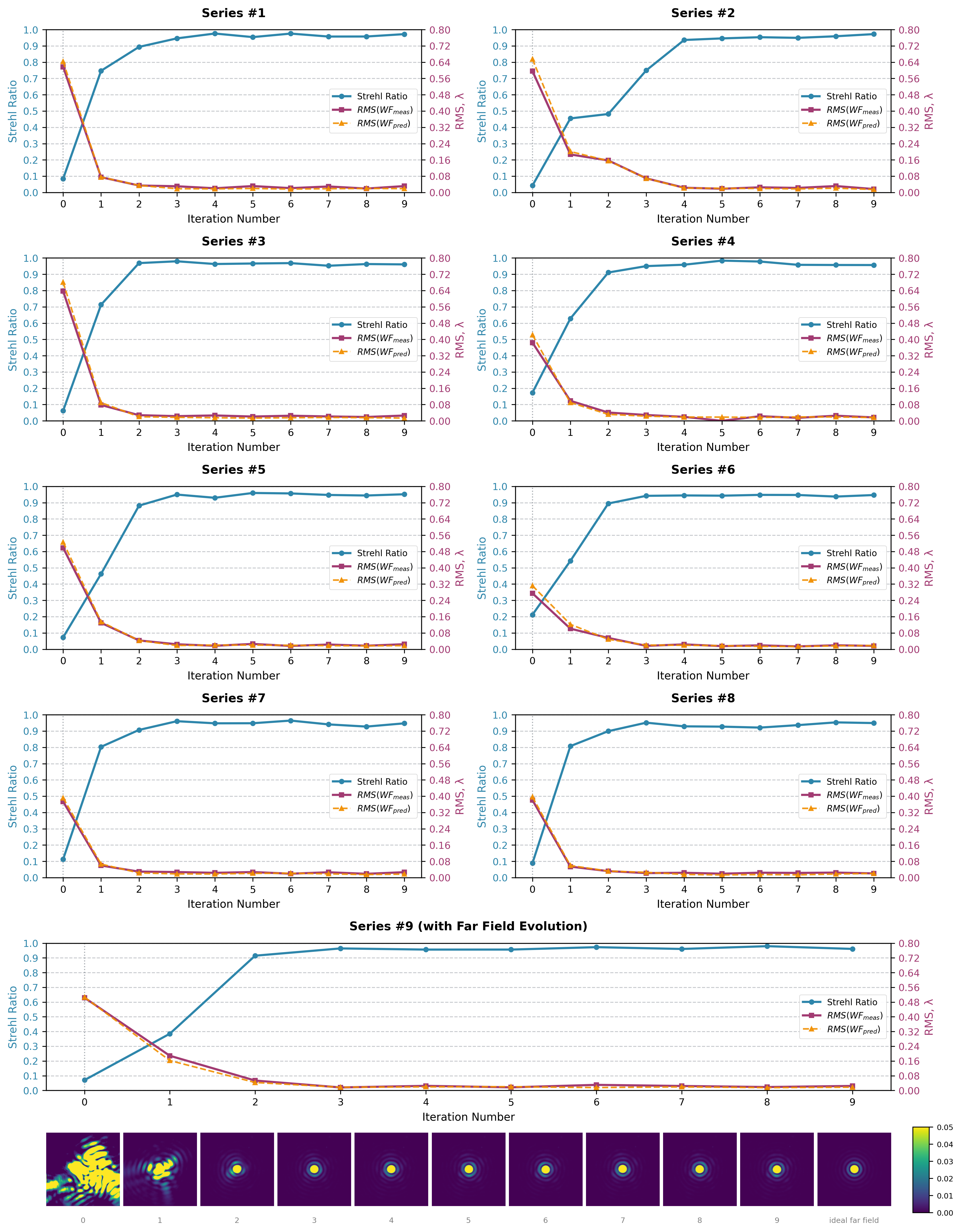}
    \caption{\label{fig:exp_res} Experimental results of the iterative wavefront correction based on the predictions of the hybrid method. The left axis indicates the Strehl ratio; the right axis shows the RMS of the wavefront distortions measured by the WFS and predicted by the hybrid method. For the final correction series, the normalized intensity distributions at the focal plane for each iteration and for the ideal focusing case are displayed, rendered with a colormap scaled from 0 to 0.05.}
\end{figure*}

The results of iterative wavefront correction based on the predictions of the hybrid method are shown in Fig.~\ref{fig:exp_res}. The left vertical axis of the plots denotes the Strehl ratio, while the right axis indicates the RMS of the wavefront distortions measured by the WFS and predicted by the hybrid method. For the final correction series, focal plane images are also displayed for each iteration and for the calculated diffraction-limited focusing case. These distributions are normalized to their respective peak values and plotted in a palette from 0 to 0.05.

The hybrid method achieves a Strehl ratio of approximately 0.96, which corresponds to a reduction of the residual RMS of the wavefront distortions to below $0.032\lambda$. The convergence of the correction loop is reached within 2 to 4 iterations. In the numerical simulations, the RMS prediction error of the wavefront distortions was on the order of $0.01\lambda$ to $0.03\lambda$. Consequently, for synthetic data, a single correction iteration is sufficient to achieve a comparable or superior performance. The discrepancy between the numerical and experimental results is attributed to several factors: WFS measurement uncertainty, noise during the acquisition of the intensity distributions at the focal and defocused planes, and the limited spatial fitting capability of the deformable mirror.

The lack of further improvement is explained not only by WFS limitations and registration noise but also by residual wavefront distortions residing in higher-order Zernike modes. The deformable mirror used in this study has a limited capacity for correcting high-spatial-frequency wavefront distortions due to the specific number and geometry of its electrodes, which enforces a 25-mode correction limit. Therefore, mitigating higher spatial frequencies would require a DM with a higher actuator density.

With the reference wavefront corresponding to the realization with a Strehl ratio of 0.98 (series 4, iteration 5), the next stage of the experiment was to measure the efficiency dependence, similar to the one obtained from the numerical data (see Figure \ref{fig:num_result}).

To evaluate the quality of the correction of wavefront distortions, an efficiency metric is defined as:
$$\text{efficiency} = 1 - \frac{\text{RMS}(\hat{W}_{\text{pred}} - W_{\text{meas}})}{\text{RMS}(W_{\text{meas}})},$$
where $W_{\text{meas}}$ represents the wavefront distortions measured by the WFS. This metric characterizes the residual wavefront distortions relative to the initial state. The measurement uncertainty, corresponding to the WFS measurement error, is $\sigma_{\text{meas}} = 0.025\lambda$. When the measured wavefront distortions are comparable to this uncertainty ($\text{RMS}(W_{\text{meas}}) \sim \sigma_{\text{meas}}$), the error contribution becomes indistinguishable from the signal, which reduces the reliability of the efficiency estimation. Consequently, at low RMS values of the measured wavefront distortions, the metric becomes statistically unstable as noise dominates. Based on this, an applicability threshold of $2\sigma_{\text{meas}} = 0.05\lambda$ was established, ensuring the error contribution does not exceed 50\%. Data points falling below this threshold were excluded from the analysis.

\begin{figure}[H]
    \centering
    \includegraphics[width=0.48\textwidth]{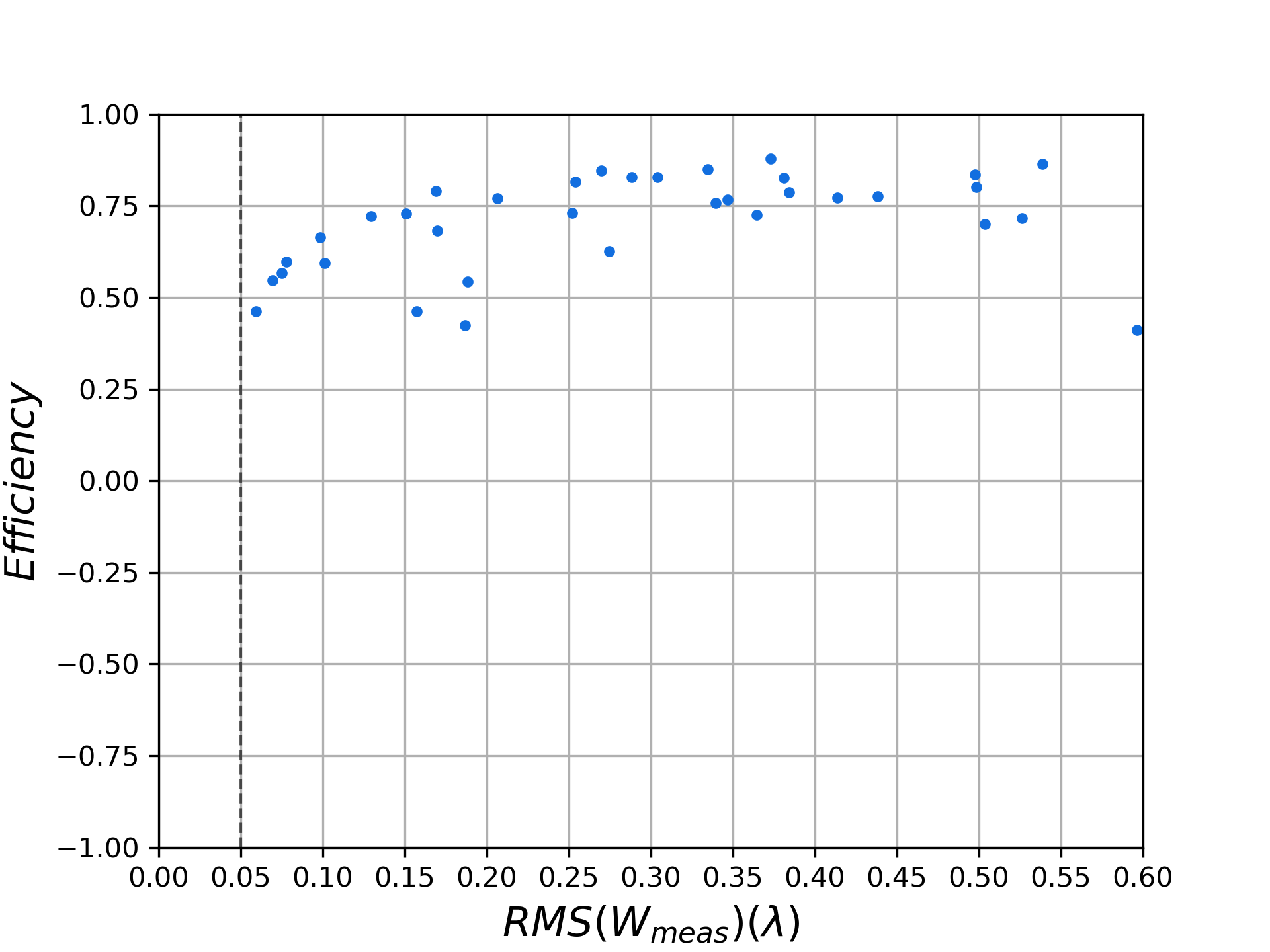}
    \caption{\label{fig:exp_res1} Efficiency of the hybrid method on experimental data as a function of the RMS of wavefront distortions.}
\end{figure}

The dependence of efficiency on the initial wavefront distortions for the experimental data is shown in Fig.~\ref{fig:exp_res1}. For an RMS of wavefront distortions exceeding $0.15\lambda$, the efficiency is approximately 75\%. Below $0.15\lambda$, the efficiency begins to decline. This decrease occurs because the metric is relative. At small initial wavefront distortions, even minor absolute deviations significantly affect the result. Furthermore, in this range, the influence of WFS measurement and registration errors is amplified. Nevertheless, the average efficiency remains above 50\%, facilitating convergence within several iterations of the hybrid method, as demonstrated in Fig.~\ref{fig:exp_res}.

\section{\label{sec:conclusion} Conclusion}

In this work, the hybrid method for the reconstruction of wavefront distortions and adaptive optics system calibration was developed, combining deep learning with the local optimization algorithm. A trained convolutional neural network predicts an initial estimate of the wavefront distortions, which is subsequently improved by the L-BFGS algorithm. When optimizing the amplitudes of the Zernike modes with the L-BFGS algorithm, three cross-sections of the intensity distribution were used: at the focal plane, and at two defocused planes with positive and negative defocus.

Numerical experiments demonstrate that the hybrid method achieves high accuracy in the reconstruction of wavefront distortions. The mean RMS reconstruction error is on the order of $10^{-2}\lambda$ across an RMS of wavefront distortions ranging from 0 to $1.3\lambda$, regardless of the number of modes in the data. The correction efficiency exceeds $0.99$ in 80\% of the cases and drops below $0.75$ in fewer than 0.5\% of the instances. The computational time is on the order of a few seconds on a CPU, demonstrating practical applicability. By contrast, relying solely on the neural network, while stable, yields an average RMS error five times higher than that of the hybrid method and fails to achieve an efficiency above $0.95$ for any combination of the magnitude of the wavefront distortions and the number of Zernike modes. Applying the L-BFGS algorithm with zero or random initialization yields efficiencies approaching unity only for small wavefront distortions. However, as the magnitude of the wavefront distortions or the number of Zernike modes increases, the efficiency decreases. Consequently, the overall accuracy is significantly lower than that achieved by the hybrid method.

Experimental validation demonstrates the practical applicability of the developed method. The hybrid method converges within 2 to 4 iterations, achieving a Strehl ratio of $0.96 \pm 0.02$ and reducing the residual RMS of the wavefront distortions to $0.032\lambda$. Discrepancies in the convergence rates between the experimental and numerical results are attributed to measurement noise, spatial scale calibration uncertainties, and inherent errors in both the sensing and correction of the wavefront distortions. Incorporating the third intensity distribution as a regularizing constraint during the optimization using the L-BFGS algorithm ensures the robustness of the method and maintains a stable correction performance post-convergence, thereby preventing error accumulation.

\begin{acknowledgments}

The authors acknowledge the support of the Ministry of Science and Higher Education of the Russian Federation, project FSWR-2026-0007 (Y.R., V.V., and I.M., development of the hybrid method, computational experiments), project FFUF–2024–0038 (A.K., K.B., S.P., A.S., physical experiments, preprocessing of experimental data). The hardware platform for the experiment was provided with the support of the Russian Science Foundation project No. 25-62-00019. The authors acknowledge the use of computational resources provided by Lobachevsky University.

\end{acknowledgments}

\section*{Data Availability Statement}

The data that support the findings of this study are openly available in https://cloud.unn.ru/s/6Z5eeTsxi9is6RA.

\nocite{*}
\bibliography{references}

@article{wyant1992basic,
  title={Basic wavefront aberration theory for optical metrology},
  author={Wyant, James C and Creath, Katherine},
  journal={Applied optics and optical engineering},
  volume={11},
  number={part 2},
  pages={28--39},
  year={1992},
  publisher={Academic New York}
}

@article{soloviev2020adaptive,
  title={Adaptive system for wavefront correction of the PEARL laser facility},
  author={Soloviev, Aleksandr Andreevich and Kotov, Aleksandr Vladimirovich and Perevalov, Sergei Evgen'evich and Esyunin, Maksim Viktorovich and Starodubtsev, Mikhail Viktorovich and Alexandrov, Aleksandr Georgievich and Galaktionov, Ilia Vladimirovich and Samarkin, Vadim Vasil'evich and Kudryashov, Aleksei Valer'evich and Ginzburg, Vladislav Naumovich and others},
  journal={Kvantovaya Elektronika},
  volume={50},
  number={12},
  pages={1115--1122},
  year={2020},
  publisher={Lebedev Physical Institute of the Russian Academy of Sciences}
}

@article{soloviev2024research,
  title={Research in plasma physics and particle acceleration using the PEARL petawatt laser},
  author={Soloviev, Aleksandr Andreevich and Burdonov, Konstantin Feliksovich and Ginzburg, Vladislav Naumovich and Glyavin, Mikhail Yur'evich and Zemskov, Roman Sergeevich and Kotov, Aleksandr Vladimirovich and Kochetkov, Anton Andreevich and Kuzmin, Aleksei Aleksandrovich and Murzanev, Aleksey Andreevich and Mukhin, Ivan Borisovich and others},
  journal={Uspekhi Fizicheskikh Nauk},
  volume={194},
  number={3},
  pages={313--335},
  year={2024},
  publisher={Russian Academy of Sciences, Branch of Physical Sciences}
}

@inproceedings{rukosuev2002adaptive,
  title={Adaptive optical system based on bimorph mirror and Shack-Hartmann wavefront sensor},
  author={Rukosuev, Alexey L and Alexandrov, Alexander and Zavalova, Valentina Ye and Samarkin, Vadim V and Kudryashov, Alexis V},
  booktitle={High-Resolution Wavefront Control: Methods, Devices, and Applications III},
  volume={4493},
  pages={261--268},
  year={2002},
  organization={SPIE}
}

@inproceedings{dean2006phase,
  title={Phase retrieval algorithm for JWST flight and testbed telescope},
  author={Dean, Bruce H and Aronstein, David L and Smith, J Scott and Shiri, Ron and Acton, D Scott},
  booktitle={Space telescopes and instrumentation I: optical, infrared, and millimeter},
  volume={6265},
  pages={314--330},
  year={2006},
  organization={SPIE}
}

@article{fienup1993hubble,
  title={Hubble Space Telescope characterized by using phase-retrieval algorithms},
  author={Fienup, James R and Marron, Joseph C and Schulz, Timothy J and Seldin, John H},
  journal={Applied optics},
  volume={32},
  number={10},
  pages={1747--1767},
  year={1993},
  publisher={Optical Society of America}
}

@article{jin2020wavefront,
  title={Wavefront reconstruction based on deep transfer learning for microscopy},
  author={Jin, Yuncheng and Chen, Jiajia and Wu, Chenxue and Chen, Zhihong and Zhang, XIngyu and Shen, Hui-liang and Gong, Wei and Si, Ke},
  journal={Optics Express},
  volume={28},
  number={14},
  pages={20738--20747},
  year={2020},
  publisher={Optical Society of America}
}

@article{booth2014adaptive,
  title={Adaptive optical microscopy: the ongoing quest for a perfect image},
  author={Booth, Martin J},
  journal={Light: Science \& Applications},
  volume={3},
  number={4},
  pages={e165--e165},
  year={2014},
  publisher={Nature Publishing Group}
}

@article{sulai2014non,
  title={Non-common path aberration correction in an adaptive optics scanning ophthalmoscope},
  author={Sulai, Yusufu N and Dubra, Alfredo},
  journal={Biomedical optics express},
  volume={5},
  number={9},
  pages={3059--3073},
  year={2014},
  publisher={Optical Society of America}
}

@inproceedings{bennet2018free,
  title={Free-space quantum communication link with adaptive optics},
  author={Bennet, F and Thearle, O and Roberts, L and Smith, J and Spollard, J and Shaddock, D and Lam, P},
  booktitle={In Proc. of 2018 Advanced Maui Optical and Space Surveillance Technologies Conference (AMOS), Hawaii},
  year={2018}
}

@article{tyson1996adaptive,
  title={Adaptive optics and ground-to-space laser communications},
  author={Tyson, Robert K},
  journal={Applied optics},
  volume={35},
  number={19},
  pages={3640--3646},
  year={1996},
  publisher={Optical Society of America}
}

@article{li2017bp,
  title={BP artificial neural network based wave front correction for sensor-less free space optics communication},
  author={Li, Zhaokun and Zhao, Xiaohui},
  journal={Optics Communications},
  volume={385},
  pages={219--228},
  year={2017},
  publisher={Elsevier}
}

@article{kotov2021adaptive,
  title={Adaptive system for correcting optical aberrations of high-power lasers with dynamic determination of the reference wavefront},
  author={Kotov, Aleksandr Vladimirovich and Perevalov, Sergei Evgen'evich and Starodubtsev, Mikhail Viktorovich and Zemskov, Roman Sergeevich and Alexandrov, Aleksandr Georgievich and Galaktionov, Ilia Vladimirovich and Kudryashov, Aleksei Valer'evich and Samarkin, Vadim Vasil'evich and Soloviev, Aleksandr Andreevich},
  journal={Quantum Electronics},
  volume={51},
  number={7},
  pages={593--596},
  year={2021},
  publisher={Kvantovaya Elektronika, Turpion Ltd and IOP Publishing}
}

@article{yoon2019achieving,
  title={Achieving the laser intensity of 5.5$\times$ 1022 W/cm2 with a wavefront-corrected multi-PW laser},
  author={Yoon, Jin Woo and Jeon, Cheonha and Shin, Junghoon and Lee, Seong Ku and Lee, Hwang Woon and Choi, Il Woo and Kim, Hyung Taek and Sung, Jae Hee and Nam, Chang Hee},
  journal={Optics express},
  volume={27},
  number={15},
  pages={20412--20420},
  year={2019},
  publisher={Optical Society of America}
}

@article{pirozhkov2017approaching,
  title={Approaching the diffraction-limited, bandwidth-limited Petawatt},
  author={Pirozhkov, Alexander S and Fukuda, Yuji and Nishiuchi, Mamiko and Kiriyama, Hiromitsu and Sagisaka, Akito and Ogura, Koichi and Mori, Michiaki and Kishimoto, Maki and Sakaki, Hironao and Dover, Nicholas P and others},
  journal={Optics express},
  volume={25},
  number={17},
  pages={20486--20501},
  year={2017},
  publisher={Optical Society of America}
}

@article{sauvage2007calibration,
  title={Calibration and precompensation of noncommon path aberrations for extreme adaptive optics},
  author={Sauvage, Jean-Fran{\c{c}}ois and Fusco, Thierry and Rousset, G{\'e}rard and Petit, Cyril},
  journal={Journal of the Optical Society of America A},
  volume={24},
  number={8},
  pages={2334--2346},
  year={2007},
  publisher={Optical Society of America}
}

@article{gerchberg1972practical,
  title={A practical algorithm for the determination of the phase from image and diffraction plane pictures},
  author={Gerchberg, Ralph W},
  journal={Optik},
  volume={35},
  number={2},
  pages={237--246},
  year={1972}
}

@article{piatrou2007beaconless,
  title={Beaconless stochastic parallel gradient descent laser beam control: numerical experiments},
  author={Piatrou, Piotr and Roggemann, Michael},
  journal={Applied optics},
  volume={46},
  number={27},
  pages={6831--6842},
  year={2007},
  publisher={Optical Society of America}
}

@article{yang2007intracavity,
  title={Intracavity transverse modes controlled by a genetic algorithm based on Zernike mode coefficients},
  author={Yang, Ping and Ao, MingWu and Liu, Yuan and Xu, Bing and Jiang, WenHan},
  journal={Optics express},
  volume={15},
  number={25},
  pages={17051--17062},
  year={2007},
  publisher={Optical Society of America}
}

@article{paine2018machine,
  title={Machine learning for improved image-based wavefront sensing},
  author={Paine, Scott W and Fienup, James R},
  journal={Optics letters},
  volume={43},
  number={6},
  pages={1235--1238},
  year={2018},
  publisher={Optical Society of America}
}

@article{orban2021focal,
  title={Focal plane wavefront sensing using machine learning: performance of convolutional neural networks compared to fundamental limits},
  author={Orban De Xivry, Gilles and Quesnel, Maxime and Vanberg, PO and Absil, Olivier and Louppe, Gilles},
  journal={Monthly Notices of the Royal Astronomical Society},
  volume={505},
  number={4},
  pages={5702--5713},
  year={2021},
  publisher={Oxford University Press}
}

@article{kotov2025retrieval,
  title={Retrieval of the Wavefront of a Laser Beam Via the Analysis of the in-Focus and Out-of-Focus Intensity Distributions with a Convolutional Neural Network},
  author={Kotov, AV and Rodimkov, Yu A and Meyerov, IB and Volokitin, VD and Perevalov, SE and Burdonov, KF and Zemskov, RS and Soloviev, A},
  journal={Radiophysics and Quantum Electronics},
  pages={1--6},
  year={2025},
  publisher={Springer}
}

@inproceedings{sheldakova2004genetic,
  title={Genetic and hill-climbing algorithms for laser beam correction},
  author={Sheldakova, Julia V and Rukosuev, Alexey L and Kudryashov, Alexis V},
  booktitle={Laser Resonators and Beam Control VII},
  volume={5333},
  pages={106--111},
  year={2004},
  organization={SPIE}
}

@inproceedings{yang200719,
  title={19-element sensorless adaptive optical system based on modified hill-climbing and genetic algorithms},
  author={Yang, Ping and Yang, Wei and Liu, Yuan and Hu, Shijie and Ao, Mingwu and Xu, Bin and Jiang, Wenham},
  booktitle={3rd International Symposium on Advanced Optical Manufacturing and Testing Technologies: Optical Test and Measurement Technology and Equipment},
  volume={6723},
  pages={36--42},
  year={2007},
  organization={SPIE}
}

@article{poland2008evaluation,
  title={Evaluation of fitness parameters used in an iterative approach to aberration correction in optical sectioning microscopy},
  author={Poland, Simon P and Wright, Amanda J and Girkin, John M},
  journal={Applied optics},
  volume={47},
  number={6},
  pages={731--736},
  year={2008},
  publisher={Optical Society of America}
}

@article{cao2014stochastic,
  title={Stochastic parallel gradient descent laser beam control algorithm for atmospheric compensation in free space optical communication},
  author={Cao, Jingtai and Zhao, Xiaohui and Li, Zhaokun and Liu, Wei and Song, Yang},
  journal={Optik},
  volume={125},
  number={20},
  pages={6142--6147},
  year={2014},
  publisher={Elsevier}
}

@article{el2005adaptive,
  title={Adaptive beam profile control using a simulated annealing algorithm},
  author={El-Agmy, R and Bulte, H and Greenaway, AH and Reid, DT},
  journal={Optics Express},
  volume={13},
  number={16},
  pages={6085--6091},
  year={2005},
  publisher={Optical Society of America}
}

@article{zommer2006simulated,
  title={Simulated annealing in ocular adaptive optics},
  author={Zommer, S and Ribak, EN and Lipson, SG and Adler, J},
  journal={Optics letters},
  volume={31},
  number={7},
  pages={939--941},
  year={2006},
  publisher={Optical Society of America}
}

@article{ke2019self,
  title={Self-learning control for wavefront sensorless adaptive optics system through deep reinforcement learning},
  author={Ke, Hu and Xu, Bing and Xu, Zhenxing and Wen, Lianghua and Yang, Ping and Wang, Shuai and Dong, Lizhi},
  journal={Optik},
  volume={178},
  pages={785--793},
  year={2019},
  publisher={Elsevier}
}

@article{landman2021self,
  title={Self-optimizing adaptive optics control with reinforcement learning for high-contrast imaging},
  author={Landman, Rico and Haffert, Sebastiaan Y and Radhakrishnan, Vikram M and Keller, Christoph U},
  journal={Journal of Astronomical Telescopes, Instruments, and Systems},
  volume={7},
  number={3},
  pages={039002--039002},
  year={2021},
  publisher={Society of Photo-Optical Instrumentation Engineers}
}

@article{liu2013hill,
  title={Hill-climbing algorithm based on Zernike modes for wavefront sensorless adaptive optics},
  author={Liu, Ying and Ma, Jianqiang and Li, Baoqing and Chu, Jiaru},
  journal={Optical Engineering},
  volume={52},
  number={1},
  pages={016601--016601},
  year={2013},
  publisher={Society of Photo-Optical Instrumentation Engineers}
}

@article{ma2011full,
  title={Full-field unsymmetrical beam shaping for decreasing and homogenizing the thermal deformation of optical element in a beam control system},
  author={Ma, Haotong and Zhou, Qiong and Xu, Xiaojun and Du, Shaojun and Liu, Zejin},
  journal={Optics Express},
  volume={19},
  number={S5},
  pages={A1037--A1050},
  year={2011},
  publisher={OSA}
}

@article{durech2021wavefront,
  title={Wavefront sensor-less adaptive optics using deep reinforcement learning},
  author={Durech, Eduard and Newberry, William and Franke, Jonas and Sarunic, Marinko V},
  journal={Biomedical optics express},
  volume={12},
  number={9},
  pages={5423--5438},
  year={2021},
  publisher={Optical Society of America}
}

@article{gutierrez2024image,
  title={Image-based wavefront correction using model-free reinforcement learning},
  author={Gutierrez, Yann and Mazoyer, Johan and Mugnier, Laurent M and Herscovici-Schiller, Olivier and Abeloos, Baptiste},
  journal={Optics Express},
  volume={32},
  number={18},
  pages={31247--31269},
  year={2024},
  publisher={Optica Publishing Group}
}

@article{gonsalves1982phase,
  title={Phase retrieval and diversity in adaptive optics},
  author={Gonsalves, Robert A},
  journal={Optical Engineering},
  volume={21},
  number={5},
  pages={829--832},
  year={1982},
  publisher={SPIE}
}

@article{fienup1993phase,
  title={Phase-retrieval algorithms for a complicated optical system},
  author={Fienup, James R},
  journal={Applied optics},
  volume={32},
  number={10},
  pages={1737--1746},
  year={1993},
  publisher={Optical Society of America}
}

@article{paxman1992joint,
  title={Joint estimation of object and aberrations by using phase diversity},
  author={Paxman, Richard G and Schulz, Timothy J and Fienup, James R},
  journal={Journal of the Optical Society of America A},
  volume={9},
  number={7},
  pages={1072--1085},
  year={1992},
  publisher={Optical Society of America}
}

@article{zhang2016hybrid,
  title={Hybrid particle swarm global optimization algorithm for phase diversity phase retrieval},
  author={Zhang, PG and Yang, CL and Xu, ZH and Cao, ZL and Mu, QQ and Xuan, L},
  journal={Optics Express},
  volume={24},
  number={22},
  pages={25704--25717},
  year={2016},
  publisher={Optical Society of America}
}

@article{zhou2021robust,
  title={Robust statistical phase-diversity method for high-accuracy wavefront sensing},
  author={Zhou, Zhisheng and Nie, Yunfeng and Fu, Qiang and Liu, Qiran and Zhang, Jingang},
  journal={Optics and Lasers in Engineering},
  volume={137},
  pages={106335},
  year={2021},
  publisher={Elsevier}
}

@article{jin2018machine,
  title={Machine learning guided rapid focusing with sensor-less aberration corrections},
  author={Jin, Yuncheng and Zhang, Yiye and Hu, Lejia and Huang, Haiyang and Xu, Qiaoqi and Zhu, Xinpei and Huang, Limeng and Zheng, Yao and Shen, Hui-Liang and Gong, Wei and others},
  journal={Optics express},
  volume={26},
  number={23},
  pages={30162--30171},
  year={2018},
  publisher={Optical Society of America}
}

@article{li2022prediction,
  title={Prediction of wavefront distortion for wavefront sensorless adaptive optics based on deep learning},
  author={Li, Yushuang and Yue, Dan and He, Yihao},
  journal={Applied Optics},
  volume={61},
  number={14},
  pages={4168--4176},
  year={2022},
  publisher={Optica Publishing Group}
}

@article{andersen2020image,
  title={Image-based wavefront sensing for astronomy using neural networks},
  author={Andersen, Torben and Owner-Petersen, Mette and Enmark, Anita},
  journal={Journal of Astronomical Telescopes, Instruments, and Systems},
  volume={6},
  number={3},
  pages={034002--034002},
  year={2020},
  publisher={Society of Photo-Optical Instrumentation Engineers}
}

@article{guo2019improved,
  title={Improved machine learning approach for wavefront sensing},
  author={Guo, Hongyang and Xu, Yangjie and Li, Qing and Du, Shengping and He, Dong and Wang, Qiang and Huang, Yongmei},
  journal={Sensors},
  volume={19},
  number={16},
  pages={3533},
  year={2019},
  publisher={MDPI}
}

@article{vanberg2019machine,
  title={Machine learning for image-based wavefront sensing},
  author={Vanberg, Pierre-Olivier and others},
  year={2019},
  publisher={Universit{\'e} de Li{\`e}ge, Li{\`e}ge, Belgique}
}

@article{gu2020algorithm,
  title={An algorithm combining convolutional neural networks with SPGD for SLAO in FSOC},
  author={Gu, Haijun and Liu, Meiqi and Liu, Haoyu and Yang, Xue and Liu, Wei},
  journal={Optics Communications},
  volume={475},
  pages={126243},
  year={2020},
  publisher={Elsevier}
}

@article{zhou2023generalization,
  title={Generalization of learned Fourier-based phase-diversity wavefront sensing},
  author={Zhou, Zhisheng and Fu, Qiang and Zhang, Jingang and Nie, Yunfeng},
  journal={Optics Express},
  volume={31},
  number={7},
  pages={11729--11744},
  year={2023},
  publisher={Optica Publishing Group}
}

@article{wang2025improved,
  title={Improved Phase Diversity Wavefront Sensing with a Deep Learning-Driven Hybrid Optimization Approach},
  author={Wang, Yangchen and Wen, Ming and Ma, Hongcai},
  booktitle={Photonics},
  volume={12},
  number={3},
  pages={235},
  year={2025},
  organization={MDPI}
}

@inproceedings{paine2018smart,
  title={Smart starting guesses from machine learning for phase retrieval},
  author={Paine, Scott W and Fienup, James R},
  booktitle={Space Telescopes and Instrumentation 2018: Optical, Infrared, and Millimeter Wave},
  volume={10698},
  pages={1681--1687},
  year={2018},
  organization={SPIE}
}

@article{liu1989limited,
  title={On the limited memory BFGS method for large scale optimization},
  author={Liu, Dong C and Nocedal, Jorge},
  journal={Mathematical programming},
  volume={45},
  number={1},
  pages={503--528},
  year={1989},
  publisher={Springer}
}

@book{nocedal2006numerical,
  title={Numerical optimization},
  author={Nocedal, Jorge and Wright, Stephen J},
  year={2006},
  publisher={Springer}
}

@article{rodimkov2026using,
  author = {Rodimkov, Y. A.},
  title = {Using circular buffer to improve accuracy of wavefront distortion reconstruction using deep learning methods for deformable mirror control problems},
  journal = {Transactions of NNSTU n.a. R.E. Alekseev},
  year = {2026},
  number = {1},
  pages = {32--42},
  doi = {10.46960/1816-210X_2026_1_32},
}

@inproceedings{ronneberger2015u,
  title={U-net: Convolutional networks for biomedical image segmentation},
  author={Ronneberger, Olaf and Fischer, Philipp and Brox, Thomas},
  booktitle={International Conference on Medical image computing and computer-assisted intervention},
  pages={234--241},
  year={2015},
  organization={Springer}
}

@article{kingma2014adam,
  title={Adam: A method for stochastic optimization},
  author={Kingma, Diederik P and Ba, Jimmy},
  journal={arXiv preprint arXiv:1412.6980},
  year={2014}
}

@misc{akaoptics,
  author = {{AKA Optics}},
  url = {https://www.akaoptics.com/},
  note = {Accessed: 2026-04-30}
}
\end{document}